\documentclass[graybox, nosecnum]{svmult}
\usepackage{mathptmx}       % selects Times Roman as basic font
\usepackage{helvet}         % selects Helvetica as sans-serif font
\usepackage{courier}        % selects Courier as typewriter font
\usepackage{type1cm}        % activate if the above 3 fonts are
\usepackage{makeidx}         % allows index generation
\usepackage{graphicx}        % standard LaTeX graphics tool
\usepackage{multicol}        % used for the two-column index
\usepackage[bottom]{footmisc}% places footnotes at page bottom
\usepackage{hyperref}        %for hyperlinks
\usepackage{soul}            % for high-lighting of text

\usepackage{amsmath, amssymb, amsfonts, bm}

\usepackage{tabularx}
\usepackage{booktabs}
\newcommand{\beq}{\begin{equation}}
\newcommand{\eeq}{\end{equation}}
\newcommand{\beqa}{\begin{eqnarray}}
\newcommand{\eeqa}{\end{eqnarray}}
\newcommand{\nn}{\nonumber \\ }

\definecolor{red1}{RGB}{230,85,13}

\makeindex             % used for the subject index
\begin{document}
%\tableofcontents{}
\title*{Semi-local nuclear forces from chiral EFT: State-of-the-art \&
challenges}
\author{Evgeny Epelbaum\thanks{corresponding author}, Hermann
  Krebs and  Patrick Reinert}
\institute{  Evgeny Epelbaum \at Institut f\"ur Theoretische Physik II, Ruhr-Universit\"at Bochum,  D-44780 Bochum,
 Germany, \email{evgeny.epelbaum@rub.de}
\and Hermann Krebs  \at Institut f\"ur Theoretische Physik II, Ruhr-Universit\"at Bochum,  D-44780 Bochum,
Germany, \email{hermann.krebs@rub.de}
\and Patrick Reinert \at Institut f\"ur Theoretische Physik II, Ruhr-Universit\"at Bochum,  D-44780 Bochum,
Germany, \email{patrick.reinert@rub.de}
}
\maketitle
\abstract{Recently, a new generation of nuclear forces has been
  developed in the framework of chiral EFT. An important feature of these
  potentials is a novel semi-local regularization approach that
  combines the advantages of a local regulator for long-range
  interactions with the convenience of an angle-independent nonlocal
  regulator for contact interactions. The authors discuss the key features of
  the semi-local two-nucleon potentials and demonstrate their
  outstanding performance in the two-nucleon sector by showing
  selected results up to fifth order in the EFT
  expansion. Also reviewed are applications to heavier systems, which are currently limited
  to third chiral order. This limitation reflects the conceptual difficulty in
  constructing a \textit{consistently regularized} many-body forces and
  current operators and affects all currently available interactions.
  The authors outline possible ways to tackle this problem and discuss
  future directions in the field. }

\section{\textit{Introduction}}

In the past decade, a large number of nuclear potentials have been
developed in the framework of chiral EFT. These interactions
differ by the choices of degrees of freedom in the effective Lagrangian, 
the orders in the EFT expansion, the employed regulators, the
values for the low-energy constants (LECs) and the strategies for
their determination, the treatment of relativistic and
isospin-breaking corrections and by many other aspects.
It is, therefore, important to start with briefly summarizing the main principles and 
the general framework used to develop the semi-local momentum-space
regularized (SMS) interactions of Refs.~\cite{Reinert:2017usi,Reinert:2020mcu}. 
\begin{itemize}
\item The expressions for nuclear forces
and current operators are derived from the heavy-baryon effective Lagrangian for  pions and
  nucleons
via a  perturbative expansion in powers of $Q
\in  \{p/\Lambda_{\rm b}, \, M_\pi / \Lambda_{\rm b} \}$.  Here,
$M_\pi$ denotes the pion mass while $p \sim M_\pi$ stands for  a typical 
three-momentum scale for low-energy few-nucleon processes under consideration.
%we are
%interested in.
The breakdown scale of the chiral EFT expansion
$\Lambda_{\rm b}$ in the two-nucleon sector is estimated to be
$\Lambda_{\rm b} \sim 650$~MeV \cite{Epelbaum:2014efa,Furnstahl:2015rha,Epelbaum:2019wvf}.    
\item
Following Weinberg \cite{Weinberg:1991um}, the nucleon mass $m$ is
treated as a heavier scale
as compared to $\Lambda_{\rm b}$, $m \sim \Lambda_{\rm b}^2/M_\pi$. 
In Tab.~\ref{tab:Forces}, various types of contributions
to the nuclear forces in this framework are listed, most of which have already been
worked out using dimensional regularization to deal with divergent
loop integrals. Notice that nuclear potentials and current operators
are not uniquely defined, and their derivation is considerably
more demanding than just calculating Feynman diagrams, see
Refs.~\cite{Epelbaum:2019kcf} for details.   
\item
The resulting nuclear potentials are regularized with a \emph{finite} cutoff
$\Lambda \sim \Lambda_{\rm b}$ \cite{Epelbaum:2009sd,Epelbaum:2018zli} that is chosen sufficiently soft to
prevent the appearance of spurious deeply bound states. The functional
form of the employed semi-local regulator will be specified in the
next section. The authors of Ref.~\cite{Reinert:2017usi}
do not allow for tuning the functional form of the
regulator in specific  partial waves to
improve the description of experimental data. The SMS interactions
discussed here are available for the cutoffs $\Lambda = 400$, $450$, $500$
and $550$~MeV.  The residual dependence of observables
on $\Lambda$ probes the impact of contact interactions beyond the
accuracy level of the calculation and is used to validate the
estimated truncation uncertainty.  Throughout this chapter, the
quoted truncation errors correspond to the Bayesian model $\bar C_{0.5-10}^{650}$ from
    Ref.~\cite{Epelbaum:2019zqc}.   
\item
For the pion-nucleon LECs, the values  from the Roy-Steiner
equation analysis of Ref.~\cite{Hoferichter:2015tha} are employed. LECs entering the NN contact
interactions are determined from neutron-proton and proton-proton
scattering data. Notice that $3$ out of $15$ LECs accompanying the
NN contact interactions at fourth order (N$^3$LO) parametrize the
off-shell dependence of the potential and therefore cannot be (reliably)
determined in the NN sector. In
Refs.~\cite{Reinert:2017usi,Reinert:2020mcu}, the authors employed a convention
to eliminate these off-shell terms via an appropriate unitary
transformation. This implies that certain linear
combinations of N$^4$LO short-range contributions to the three-nucleon force
(3NF) and the NN charge density operator are enhanced and appear
already at N$^3$LO \cite{Girlanda:2020pqn}. This convention results in
$21$ isospin-invariant
NN contact interactions in total at N$^3$LO and N$^4$LO, which can all
be reliably determined from the neutron-proton and proton-proton
experimental data as will be described below. Therefore, there seems
to be no need to employ heavier nuclei to extract the corresponding
LECs as done e.g.~in Ref.~\cite{Ekstrom:2015rta}. Similarly, the LECs entering the
3NF at N$^2$LO are determined from 3N data. 
\item
All charge-independence and charge-symmetry breaking contributions
to the NN force up to N$^4$LO were included in the updated version of
the original SMS potentials \cite{Reinert:2017usi} described in Ref.~\cite{Reinert:2020mcu}. 
\end{itemize}

Below, some of the key features of SMS NN potentials of
Refs.~\cite{Reinert:2017usi,Reinert:2020mcu} will be reviewed. In
particular, the semi-local regularization
approach and the partial wave analysis (PWA) of NN scattering
data using the SMS chiral NN potentials will be discussed.
The results for
phase shifts and NN observables will be compared with 
alternative PWAs and with different chiral EFT potentials. The authors also discuss selected
applications in the NN sector and for heavier systems and outline
ongoing efforts towards developing \emph{consistent} 3NFs and current
operators beyond N$^2$LO.

%This chapter is organized as follows. In section \ref{sec2}, we
%discuss the semi-local regularization approach and review the partial
%wave analysis (PWA) of NN scattering data using the SMS chiral
%interactions. We compare our results with alternative PWA and chiral
%EFT potentials and present selected applications in the NN sector.

%
\newcolumntype{Y}{>{\raggedright\arraybackslash}X}
\newcolumntype{W}{>{\raggedleft\arraybackslash}X}
\newcolumntype{Z}{>{\centering\arraybackslash}X}
\begin{table}[t]
% \begin{tabularx}{\textwidth}{YWWWWW}
\begin{tabularx}{\textwidth}{>{\hsize=0.35\hsize}Y>{\hsize=0.9\hsize}Z>{\hsize=0.9\hsize}Z>{\hsize=1.1\hsize}Z>{\hsize=1.375\hsize}Z>{\hsize=1.375\hsize}Z}  
  \toprule
& LO ($Q^0$) & NLO ($Q^2$) & N$^2$LO ($Q^3$) & N$^3$LO ($Q^4$) & N$^4$LO ($Q^5$)  \\\midrule
2NF & $1\pi$, $\; {\rm NN}_{{\color{red1}[2]}}$ & $2\pi$, $\; {\rm
 NN}_{{\color{red1}[7]}}$ & ${2\pi}$& $\; {2\pi}$,  $\; {3\pi}$, ${\rm
  NN}_{{\color{red1}[15]}}$$^{\rm a}$& ${2\pi}$,  $\; {3\pi}$ \\[3pt]\midrule
3NF & --- & --- & $2\pi$, $ 1\pi$-$ {\rm 
  NN}_{{\color{red1}[1]}}$,
$ {\rm  NNN}_{{\color{red1}[1]}}^{\phantom{[1]}}$ & $2\pi$, $\; 1\pi $-$ 2\pi$, $_{\phantom{[}} {\rm
  ring}$,
$1 \pi$-${\rm  NN}$,
$\; 2\pi$-${\rm  NN}_{\phantom{[}}^{\phantom{[}}$ &
${2\pi}$, $\; 1\pi $-$
  2\pi$, $_{\phantom{[}}^{\phantom{[}} {\rm ring}$,  $^{\phantom{[}} 1\pi$-${\rm 
  NN}_{{\color{red1}[{\rm unknown}]}}$$^{\rm b}$,
$2\pi $-${\rm NN}$$^{\rm b}$,  $\; {\rm NNN}_{{\color{red1}[13]}}^{\phantom{[}}$
\\
&&&&&\\
[-7pt] \midrule
4NF & --- & --- & --- & $^{\phantom{[}}_{\phantom{[}}{3\pi}$,  $\; {4\pi}$,
$\; 2\pi $-${\rm NN}$,  $1\pi $-${\rm NN}$-${\rm NN}^{\phantom{[}}$&
--- $^{\rm c}$\\
&&&&&\\[-9pt] 
\bottomrule
\end{tabularx}
\caption{\label{tab:Forces} Types of contributions to the nuclear
  forces at various orders in the EFT expansion using Weinberg's
  power counting. $1\pi$, $2\pi$, $3\pi$ and $4\pi$ denote one-, two-,
  three- and four-pion exchange diagrams, respectively, while NN and NNN refer to the 
  two- and three-nucleon contact interactions.  For the relativistic corrections, the assignment $m \sim
  \Lambda_b^2/M_\pi$ is made. The subscripts 
  in the square brackets indicate the numbers of LECs accompanying the
  two- and three-nucleon contact interactions. Isospin-violating
  interactions are not shown. }
   \begin{footnotesize}
    \begin{itemize}
   \item[$^{\rm a}$]3 out of $15$ operators do not contribute to the NN
     $S$-matrix in the Born approximation. The cor\-res\-ponding LECs
     can therefore not be (reliably) determined from NN data at this order.
    \item[$^{\rm b}$]These topologies have not been worked out yet.
    \item[$^{\rm c}$]These contributions have not been worked out yet.
             \end{itemize}
    \end{footnotesize}
\end{table}

\section{\textit{SMS two-nucleon potentials up to N$^4$LO$^+$}}

\subsection{\textit{Regularization and subtractions}}

Semi-locally regularized nuclear potentials up to N$^3$LO
were originally introduced in Ref.~\cite{Epelbaum:2014efa}
and extended to fifth order (N$^4$LO) in Ref.~\cite{Epelbaum:2014sza}.
The term "semi-local" refers to a local regularization method for
the long-range interactions mediated by the exchange of a single or multiple
pions in combination with a nonlocal (angle-independent) cutoff for
contact terms. In the original papers \cite{Epelbaum:2014efa,Epelbaum:2014sza}, the
local regulator was implemented in coordinate space. This somewhat
{\it ad hoc} procedure was replaced by a more thorough momentum-space approach 
in Ref.~\cite{Reinert:2017usi}. Throughout this chapter, the focus is on the SMS
2N potentials introduced in that paper and further developed in
Ref.~\cite{Reinert:2020mcu}, as well as on selected applications involving 
the 3NF regularized using the same approach.  

The main motivation to employ a local regulator for long range
interactions is to avoid the appearance of long-range regulator
artefacts. Consider, for example, the $1\pi$-exchange potential $V^{1 \pi } (q)
= \alpha /(q^2 + M_\pi^2)$ with $\vec q = \vec p ' - \vec p$ being the
momentum transfer, $\vec p$ and $\vec p '$ the initial and final
momenta of the nucleons and $\alpha$ denoting the
spin-momentum-isospin structure. Using local and nonlocal
Gaussian-type regulators one obtains 
\beqa
V^{1 \pi }_{\Lambda, \, \rm local} (\vec q) &=& \frac{\alpha}{q^2 + M_\pi^2}
e^{- \frac{q^2+ M_\pi^2}{\Lambda^2}} = \frac{\alpha}{q^2 + M_\pi^2} -
\frac{\alpha}{\Lambda^2} + \frac{\alpha}{2 \Lambda^4} (q^2 + M_\pi^2)
+ \mathcal{O} (\Lambda^{-6})\,,\nn
V^{1 \pi }_{\Lambda, \, \rm nonlocal} (\vec q) &=&\frac{\alpha}{q^2 + M_\pi^2}
e^{- \frac{p^2 + {p'} ^2}{\Lambda^2}} = \frac{\alpha}{q^2 + M_\pi^2} -
\frac{\alpha}{\Lambda^2} \frac{p^2 + {p'} ^2}{q^2 + M_\pi^2} + \mathcal{O} (\Lambda^{-4})\,,
\eeqa
where the last equalities
%in the first and second lines 
hold for
momenta below the cutoff $\Lambda$. 
The nonlocal regulator in the second line obviously affects the analytic 
 structure of the potential by changing the residue of $V^{1 \pi } (q)$ at $q^2 = -
M_\pi^2$ and induces long-range finite-$\Lambda$ artefacts that need
to be systematically taken care of at higher
orders \cite{Gasparyan:2021edy}.  In contrast, the local regulator in the first
line preserves the analytic structure of $V^{1 \pi } (q)$, and all finite-$\Lambda$ artefacts have the form of contact
interactions which are anyway present in the potential. This feature
becomes particularly important when using soft cutoff values.  

Local regulators can, in
principle, be applied to contact interactions as well
\cite{Gezerlis:2013ipa}. This then allows   
one to reduce the degree of nonlocality of the interactions, a
particularly welcome feature for certain {\it ab initio} methods like
e.g.~the Quantum Monte Carlo technique. On the other hand, locally
regularized contact terms cannot be formed into linear
combinations that contribute to specific partial waves only, i.e.~the
one-to-one correspondence between the contact interactions and the
partial waves as given by Eq.~(A.2) of Ref.~\cite{Reinert:2017usi} is lost. This feature
significantly complicates the determination of the corresponding
LECs. A semi-local regulator allows one to preserve the analytic
structure of the long-range interactions while at the same time 
keeping the simplicity of a non-local regulator $e^{- \frac{p^2 + {p'}^2}{\Lambda^2}}$ for
contact interactions. 

The authors are now in the position to specify the form of the employed
semi-local regulator. For the (isospin invariant part of the)
$1\pi$-exchange the authors of
Refs.~\cite{Reinert:2017usi,Reinert:2020mcu}  use 
\beq
\label{temp0}
V^{1\pi}_{\Lambda} (\vec q ) = - \frac{g_A^2}{4
  F_\pi^2} \bm \tau_1 \cdot \bm \tau_2 \bigg( \frac{\vec \sigma_1 \cdot
  \vec q \, \vec \sigma_2 \cdot \vec q}{q^2 + M_\pi^2}    + C \, \vec
\sigma_1 \cdot \vec \sigma_2 \bigg)
e^{- \frac{q^2 + M_\pi^2}{\Lambda^2}}\,,
\eeq
where $\vec \sigma_i$ ($\bm \tau_i$) denote the Pauli spin (isospin) matrices
of the nucleon $i$ while 
$g_A$ and  $F_\pi$ 
are the axial-vector coupling constant of the nucleon and the  pion decay
constant, respectively. Here, the freedom to
include in the definition of $V^{1\pi}_{\Lambda} (\vec q )$ a (locally
regularized) LO contact interaction is exploited to ensure that the
Fourier transform of the resulting spin-spin potential vanishes at $r
= 0$. This fixes the subtraction constant $C$ to  
\beq
\label{subtr}
C = -\frac{\Lambda \left(\Lambda ^2-2 M_\pi^2 \right) + 2 \sqrt{\pi } M_\pi^3 e^{\frac{M_\pi^2}{\Lambda ^2}}
  {\rm erfc}\left(\frac{M_\pi}{\Lambda }\right)}{3 \Lambda ^3} \,,
 \eeq
 where ${\rm erfc} (x)$ is the complementary error function.

\begin{figure}[tb]
  %\vskip 1 true cm
  \begin{center}
    \includegraphics[width=0.7\textwidth,keepaspectratio,angle=0,clip]{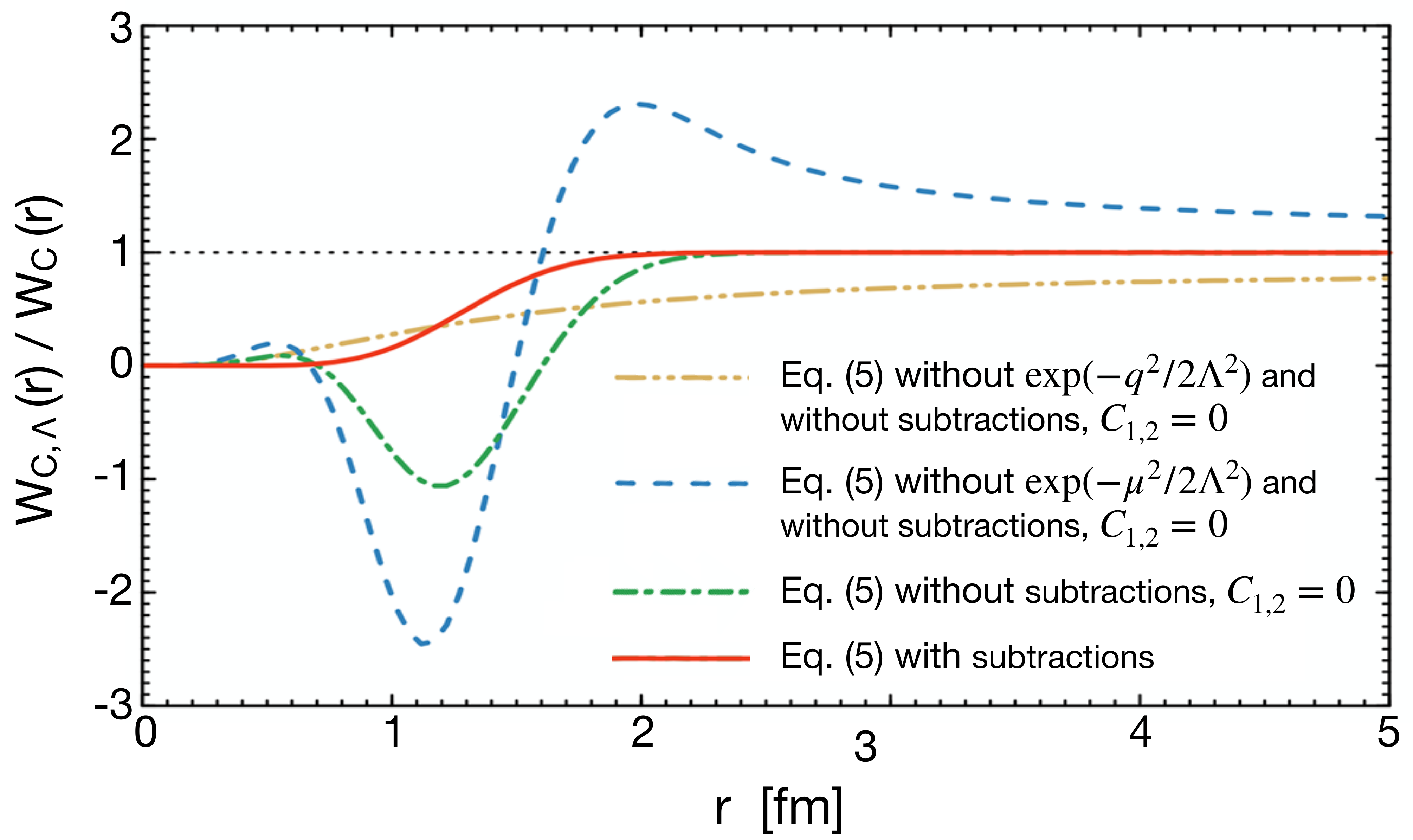}
  \end{center}
  \caption{The ratio of  the regularized to unregularized
potential, $W_{C, \, \Lambda} (r)/ W_{C} (r)$, as a function of the relative distance between the
nucleons for different regulator
choices in Eq.~(\ref{WC4}). The SMS potentials of Ref.~\cite{Reinert:2017usi,Reinert:2020mcu}
correspond to the red solid line. 
    \label{fig:regulator}
  }
\end{figure}

 For the
$2\pi$-exchange, the regulator can be easily implemented
using the spectral representation. For example, the unregularized
expression for the central $2\pi$-exchange potential at NLO, $V^{2
  \pi}  = \bm \tau_1 \cdot \bm \tau_2 W_{C} (q) + \ldots $, can be
written as a spectral integral 
\beq
\label{WC3}
W_{C} (q) =
\frac{2}{\pi} \int_{2 M_\pi}^\infty \, \frac{d \mu}{\mu^3}
\eta_{C} (\mu) \, \frac{q^4} 
{\mu^2 + q^2} \,,
\eeq
with the spectral function given by \cite{Kaiser:1997mw}
\begin{displaymath}
\eta_{C}  (\mu) = 
 \frac{\sqrt{\mu^2 - 4 M_\pi^2}}{768 \pi F_\pi^4 \mu}\,
 \, \biggl[4M_\pi^2 (5g_A^4 - 4g_A^2 -1) 
- \mu^2(23g_A^4 - 10g_A^2 -1)
+ \frac{48 g_A^4 M_\pi^4}{4 M_\pi^2 - \mu^2} \biggr] . 
\end{displaymath}
The corresponding locally regularized expression used in the
SMS potentials of Refs.~\cite{Reinert:2017usi,Reinert:2020mcu} has the form 
\beq
\label{WC4}
W_{C, \, \Lambda} (q) = e^{-\frac{q^2}{2 \Lambda^2}} \;
\frac{2}{\pi} \int_{2 M_\pi}^\infty \, \frac{d \mu}{\mu^3}
\eta_{C} (\mu) \bigg[\frac{q^4} 
{\mu^2 + q^2} + C_{1} (\mu ) +
C_{2} (\mu ) \, q^2\bigg] 
\, e^{- \frac{\mu^2}{2
    \Lambda^2}} \,.
\eeq
Again, the short-range subtraction terms are chosen to minimize the
admixtures of short-range interactions in the regularized
$2\pi$-exchange potential by enforcing
$W_{C, \, \Lambda} (r) \big|_{r=0} = 0 $ and $\frac{d^2}{dr^2} W_{C,
  \, \Lambda} (r) \big|_{r=0} = 0 $,
which leads to 
\beqa
C_{1}(\mu) &=&\frac{2 \Lambda \mu^2 \left(2 \Lambda ^4-4 \Lambda ^2 \mu ^2-
   \mu ^4 \right) + \sqrt{2 \pi } \mu ^5 e^{\frac{\mu ^2}{2 \Lambda ^2}} \left(5 \Lambda ^2+\mu ^2\right)
   {\rm erfc}\left(\frac{\mu }{\sqrt{2} \Lambda }\right)}{4 \Lambda ^5}
\,, \nn
C_{2}(\mu) &=&- \frac{2 \Lambda \left( 6 \Lambda ^6-2 \Lambda ^2 \mu ^4-  
   \mu ^6 \right) +\sqrt{2 \pi } \mu ^5 e^{\frac{\mu ^2}{2 \Lambda ^2}} \left(3 \Lambda ^2+\mu ^2\right)
   {\rm erfc}\left(\frac{\mu }{\sqrt{2} \Lambda }\right)}{12 \Lambda ^7}\,. \nonumber
\eeqa
To illustrate the effect of the regulator and to demonstrate the
importance of maintaining the analytic structure of the interaction,
Fig.~\ref{fig:regulator} shows the ratio of the regularized to unregularized
potential for different regulator choices. 
Notice that keeping either only the spectral regulator 
$e^{-\frac{\mu^2}{2 \Lambda^2}}$ or the momentum-transfer cutoff 
$e^{-\frac{q^2}{2 \Lambda^2}}$ affects the discontinuity across the
left-hand cut and results in strong distortions of the potential that
extend to large distances as shown in the figure. The difference
between the green dashed-dotted line and red solid line representing 
the actual form of the SMS regulator visualizes the
impact of the short-range subtraction terms $\propto C_{1,2} (\mu )$. 
Regularization of other long-range contributions is performed
analogously, see Ref.~\cite{Reinert:2017usi} for details and the
explicit expressions for the potential up to N$^4$LO$^+$. Here and in
what follows, the ``$^+$'' signifies the inclusion of $4$ contact
interactions from N$^5$LO in the partial waves $^1$F$_3$,   $^3$F$_3$,  
$^3$F$_2$ and $^3$F$_4$, which is necessary for performing a partial wave
analysis of proton-proton data \cite{Reinert:2017usi}. Notice that the
corresponding LECs are of natural size and {\it not} enhanced.

\subsection{\textit{Partial wave analysis of NN scattering}}

% \begin{itemize}
% \item[--]  The strategy to determine LECs of contact interactions
% \item[--]  Partial wave analysis of NN data from the PRL/Patrick's
%   thesis
% \item[--]  Resulting values for $\chi^2$
% \item[--]  Resulting values of the LECs in a plot (to emphasize naturalness)
% \end{itemize}

% {\color{red} PR, 4-5 pages. I would focus here mainly on np and pp interactions}

% \notes{phase shift differences for np, $\chi^2$ values of other potentials for combined np+pp}

Once the chiral interaction has been derived and regularized, the
numerical values of the LECs entering the potential need to be
determined from experimental data. Whenever possible,  such
LECs are fixed from the simplest process they are contributing to. As already
mentioned, the authors of
Refs.~\cite{Reinert:2017usi,Reinert:2020mcu}
use the subleading $\pi$N LECs which enter the
two-pion exchange (TPE) starting at
N$^2$LO from the recent
Roy-Steiner equation analysis of $\pi$N scattering data in
Ref.~\cite{Hoferichter:2015tha}. The isospin-invariant two-pion exchange potential is thus
parameter-free in the two-nucleon (2N) system.

The corresponding isospin-breaking corrections to the
pion-exchange potentials are also included, which are mostly parameter-free except for
the now appearing charge-dependence of the leading $\pi$N coupling
constant in the one-pion exchange and a particular contribution to
the TPE at N$^4$LO. For these terms, one has to distinguish between
three different coupling constants $f_{\pi^0\mathrm{pp}}$, $f_
{\pi^0\mathrm{nn}}$ and $f_{\pi^\pm\mathrm{pn}}$ for the interactions
between protons and neutrons with neutral or charged pions,
respectively. Extractions from $\pi$N data are only available for $f_
{\pi^\pm\mathrm{pn}}$ and therefore the authors employ their own
determination \cite{Reinert:2020mcu} of all three coupling constants
from neutron-proton and proton-proton data using the N$^4$LO$^+$
interaction discussed here.

It remains to determine the contact interaction LECs in the
short-range part of the potential. As has become customary for
higher-order chiral interactions in recent years, the authors of
Refs.~\cite{Reinert:2017usi,Reinert:2020mcu} also fit them
directly to neutron-proton and proton-proton scattering data. Many
elements of this determination are shared with the aforementioned
extraction of the charged-dependent $\pi$N coupling constants of
Ref.~\cite{Reinert:2020mcu}, but in contrast the latter uses a
Bayesian approach to integrate the cutoff $\Lambda$ and the contact
interaction LECs out of the probability density to obtain a unique
value for each of the coupling constant. Once the $\pi$N coupling
constants have been fixed, the authors choose a set of fixed values of
$\Lambda$ and adjust the contact LECs for each of them to arrive at a
fully specified set of parameters. In addition to the scattering
data, the exact reproduction of the deuteron binding
energy $B_d = 2.224575(9)$~MeV \cite{VanDerLeun:1982bhg} and of the
coherent neutron-proton scattering length $b_\mathrm{np} = -3.7405(9)$~fm
\cite{Schoen:2003my} is imposed as constraints on the fit.

The required machinery for a high-precision fit of the 2N
scattering data has been worked out by the Nijmegen group,
culminating in their seminal 1993 PWA of
Ref.~\cite{Stoks:1993tb}. One important element for the accurate
description of the scattering observables is the inclusion of the
appropriate long-range electromagnetic interactions, especially at
low energies and/or small forward angles (and for proton-proton
scattering also large backward angles). The treatment of
electromagnetic interactions by the Nijmegen group in
Ref.~\cite{Stoks:1993tb} has become {\it de facto} standard when
calculating scattering observables, and it is employed in the presented analysis
as well. The second element required for a statistically satisfactory description
of 2N scattering data is the removal of data sets that are not
compatible with the bulk of the database. The so-called
$3\sigma$-criterion established by the Nijmegen group for rejecting
such outlier data has been employed in nearly all subsequent PWAs.

In order to perform a reliable statistical testing of the data sets,
the employed nuclear interaction has to be able to achieve a
near-perfect description of the mutually consistent data. The results
for the N$^4$LO$^+$ interaction of the first version of the SMS
interaction in Ref.~\cite{Reinert:2017usi} showed that a
$\chi^2$/datum $\sim 1$ description of the data could indeed be achieved and
encouraged the authors to subsequently perform their own data selection to
arrive at a database of mutually consistent scattering data. Such a
data selection has been performed in the energy range of $E_\mathrm{lab}
= 0$--$300$~MeV in Ref.~\cite{Reinert:2020mcu} where it is
compared against the database of the recent 2013 Granada PWA of
Ref.~\cite{NavarroPerez:2013mvd}. A complete listing can be found in
Ref.~\cite{Reinert:2022thesis}. The same database is also used in the
fits of the contact interactions discussed here, but the energy range
is slightly lowered below the pion-production threshold to
$E_\mathrm{lab} = 280$~MeV.

\begin{table}[tb]
    \centering
    \smallskip

    \begin{tabular*}{\textwidth}{@{\extracolsep{\fill}}rcccc}
        \toprule
        $E_\mathrm{lab}$ bin & $\Lambda = 400$ MeV & $\Lambda = 450$ MeV & $\Lambda = 500$ MeV & $\Lambda = 550$ MeV \\
        \midrule
        \multicolumn{5}{l}{neutron-proton scattering data} \\[1pt]
        $0-100$ & 1.069 & 1.061 & 1.060 & 1.062 \\
        $0-200$ & 1.085 & 1.074 & 1.069 & 1.075 \\
        $0-280$ & 1.113 & 1.060 & 1.048 & 1.055 \\[0pt]
        \midrule
        \multicolumn{5}{l}{proton-proton scattering data} \\[1pt]
        $0-100$ & 0.876 & 0.860 & 0.866 & 0.875 \\
        $0-200$ & 0.933 & 0.909 & 0.918 & 0.942 \\
        $0-280$ & 0.956 & 0.932 & 0.950 & 0.989 \\[0pt]
        \bottomrule
    \end{tabular*}
        \caption{$\chi^2$/datum values of the N$^4$LO$^+$ potential
         for the description of neutron-proton and proton-data and
         for all considered values of the cutoff $\Lambda$. The
         energy bin for $E_\mathrm{lab} = 0-280$~MeV corresponds to
         the fitting energy range.
        \label{tab:chi2-SMS-N4LO+}}
\end{table}

Tab.~\ref{tab:chi2-SMS-N4LO+} shows the $\chi^2$/datum values for the
description of the neutron-proton and proton-proton database of the
fitted SMS N$^4$LO$^+$ interaction for all considered values of the
cutoff $\Lambda$. The database up to $E_\mathrm{lab} = 280$~MeV
consists of 2845 individual neutron-proton and 2081 individual
proton-proton data points including estimated data set
normalizations. For details regarding the definition of the $\chi^2$
measure and the estimation of normalizations, see Ref.~\cite
{Reinert:2017usi}. The excellent description of the scattering data
(especially for the cutoffs $\Lambda = 450$ and $500$~MeV) and the
independently performed data selection qualify these results to be
regarded as a PWA of 2N scattering.

The focus of a PWA is the accurate determination of the phase shifts
and mixing angles which parametrize the on-shell scattering
amplitude. Indeed, the modern determinations of the phase shift vary
only by a small amount relative to their absolute sizes and are
therefore well-known. This is especially true for the proton-proton
phase shifts where the precise scattering data constrain
the phase shifts very well, and all recent PWAs agree well with
each other. The authors therefore focus below on the results for
neutron-proton phase shifts, where the lower precision of the
neutron-proton data compared to the proton-proton data and the
different assumptions about isospin-breaking (IB) effects in the nuclear interactions
lead to a greater variation in the phase shifts.

\begin{figure}[tbp]
  %\vskip 1 true cm
  \begin{center}
    \includegraphics[width=\textwidth,keepaspectratio,angle=0,clip]{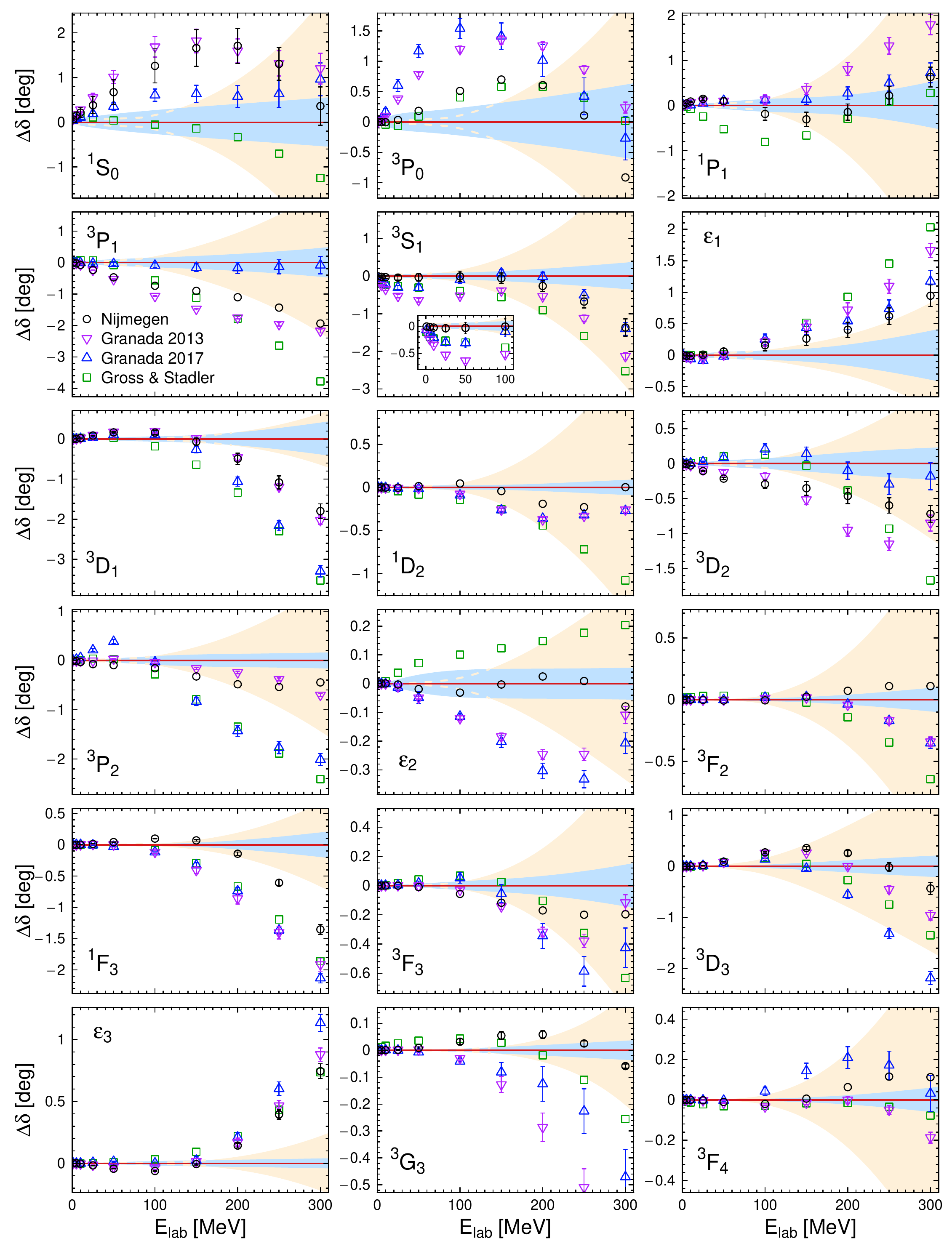}
  \end{center}
  \caption{Differences of the neutron-proton phase shifts of selected
    partial-wave analyses to the chiral SMS N$^4$LO$^+$ phase shifts of
    this work for $\Lambda = 450$~MeV. Black circles, purple down
    triangles, blue up triangles and green squares denote the results
    of the Nijmegen 1993 \cite{Stoks:1993tb}, the Granada 2013
    \cite{NavarroPerez:2013mvd}, the Granada 2017 \cite{NavarroPerez:2016eli}
    and the Gross-Stadler \cite{Gross:2008ps} PWAs, respectively, and
    the corresponding error bars denote their statistical uncertainties
    (if provided by the original publication). The peach- and light
    blue-colored bands show the truncation uncertainty
    %estimated using
  %  the Bayesian model $\bar C_{0.5-10}^{650}$ from
  %  Ref.~\cite{Epelbaum:2019zqc}
    and the combined
    statistical uncertainties of the N$^4$LO$^+$ result due to the
    NN and $\pi$N LECs, respectively.
    \label{fig:phase-shift-diff-np}
  }
\end{figure}

In order to better visually inspect the differences between selected
recent determinations of the neutron-proton phase shifts, Fig.~\ref
{fig:phase-shift-diff-np} shows their difference to the SMS N$^4$LO$^+$
results for the most accurate cutoff $\Lambda = 450$~MeV. In
particular, the authors compare to the results of the Nijmegen 1993 \cite
{Stoks:1993tb}, Gross \& Stadler 2008 \cite{Gross:2008ps}, Granada
2013 \cite{NavarroPerez:2013mvd} and Granada 2017 \cite
{NavarroPerez:2016eli} PWAs. Also shown is a Bayesian estimation of the uncertainty due to the
truncation of the chiral expansion
%(to be further discussed in the
    %     next section)
along with the combined statistical uncertainties of all
parameters. The latter are dominated by the uncertainties from the
nucleon-nucleon system while the errors of the $\pi$N LECs of
Ref.~\cite{Hoferichter:2015tha} are small. For details regarding
the employed uncertainty quantification see Ref.~\cite{Reinert:2022thesis}.

The uncertainties of the phase shifts at low energies up to $E_\mathrm
{lab} \sim 100$--$150$~MeV are dominated by the statistical errors,
whereas the truncation uncertainty becomes dominant at higher
energies. Even larger, however, is the variation between the
considered PWAs in many cases. In particular, consider the S-
and P-waves. There are some differences in the assumptions about
isospin-breaking between the considered analyses: The Nijmegen and
Granada 2013 analyses allow for a charge-dependent short-range
interaction only in the $^1S_0$ channel, whereas isovector P- and higher partial
waves only take into account the pion mass difference in the 
$1\pi$-exchange. In contrast, the analysis described above and the Granada 2017 one allow
for short-range charge dependence in both S- and isovector P-waves.
The determination from neutron-proton data is also reflected in the
statistical uncertainties of the $^1S_0$ and the isovector P-wave
phase shifts, which are 2--3 times larger than the corresponding
proton-proton phase shift uncertainties. Lastly, the Gross-Stadler
PWA is fitted to neutron-proton data only.

One significant change upon introduction of the additional short-range
IB in P-waves can be seen in the $^3P_1$ channel, where the analysis
by the authors and the
Granada 2017 one find a phase shift that is up to $2^\circ$ smaller
in magnitude than for the Nijmegen and Granada 2013 PWAs. The impact
of the IB effects is also supported by the fact that the fits the
authors performed without
the additional P-wave charge dependence are in good agreement with
the latter. However, the behavior of the corresponding Gross-Stadler
phase shift, which is completely determined by neutron-proton data,
is puzzling. The situation is less clear in other P-waves, where
there is e.g. notable variation in the maximum of the $^3P_0$ phase
shift around $E_\mathrm{lab} \sim 50$~MeV. Based on the statistical
error, the Granada 2017 result at that energy constitutes a $7\sigma$
deviation from the result of the present analysis. The authors also found statistically significant
differences in the low-energy behavior of the $^3S_1$ phase shift.
The obtained phase shift is in very good agreement with the Nijmegen analysis,
whereas the other PWAs obtain slightly smaller phase shifts in the
range of $E_\mathrm{lab} = 0$--$100$~MeV. For the Granada 2013
analysis, this amounts to a $11\sigma$ deviation from the result
obtained by the authors at
$E_\mathrm{lab} = 25$~MeV, which presumably also manifests itself in a
different result for the deuteron asymptotic S-state normalization
$A_S = 0.8829$~fm$^{-1/2}$ \cite{NavarroPerez:2013usk} compared to
other analyses.
 
% $^3S_1$: Granada 2017 closer, but still $6\sigma$

Regarding higher partial-waves in Fig.~\ref{fig:phase-shift-diff-np},
it is worth pointing out that D- and F-waves are parametrized with one
isospin-invariant short-range LEC each. The results for $^3G_3$ and
$\epsilon_3$, however, are predictions based on the long-range
potential alone. While the deviation of $\sim 1^\circ$ of the mixing
angle $\epsilon_3$ may appear large, it should be noted that the
mixing angle itself is comparatively large and reaches
$\sim 6$--$7^\circ$ at $E_\mathrm{lab} = 300$~MeV.

% also enabled extraction of charged-dependent $\pi$N coupling constants
%*EE  new table
\begin{table}[tb]
    \smallskip
    \begin{tabular*}{\textwidth}{@{\extracolsep{\fill}}llccccccc}
        \toprule
        \noalign{}
         &                   & \multicolumn{3}{c}{EMN} & \multicolumn{4}{c}{SMS} \\
        \cmidrule(lr){3-5}
        \cmidrule(lr){6-9}
         & $E_{\rm lab}$ bin & 450 MeV & 500 MeV & 550 MeV & 400 MeV & 450 MeV & 500 MeV & 550 MeV
%        \smallskip
        \\
         \midrule
        %\smallskip
                    & 0--100 & 1.302 & 1.113 & 1.235 & 1.008 & 1.021 & 1.070 & 1.140 \\
        N$^3$LO     & 0--200 & 1.549 & 1.284 & 1.426 & 1.182 & 1.353 & 1.595 & 1.904 \\
                    & 0--300$^{\rm a}$ & 2.354 & 1.503 & 1.691 & 1.601 & 2.524 & 3.903 & 5.831 \\ [0pt]
        \midrule
        %\smallskip
                    & 0--100 & 1.156 & 1.084 & 1.140 & 1.001 & 0.990 & 0.991 & 0.996 \\
        N$^4$LO$^+$ & 0--200 & 1.219 & 1.136 & 1.238 & 1.023 & 1.007 & 1.008 & 1.021 \\
                    & 0--300 & 2.019 & 1.203 & 1.315 & 1.063 & 1.013 & 1.015 & 1.042 \\ [0pt]
        %\smallskip
        \bottomrule
    \end{tabular*}
    \caption{$\chi^2$/datum of the most recent higher order chiral
     potentials for the description of the combined neutron-proton
     and proton-proton scattering data up to $E_\mathrm{lab} =
     300$~MeV. The $\chi^2$/datum values are shown for the orders
     N$^3$LO and N$^4$LO$^+$ and for all available cutoff values of
     both the SMS interaction of this work and the EMN potentials of
     Ref.~\cite{Entem:2017gor}.
     \label{tab:chi2-emn-sms}}
   \begin{footnotesize}
 \vspace{-0.2cm}
    \begin{itemize}
   \item[$^{\rm a}$] The SMS N$^3$LO potentials are fitted to
     the scattering data up to $E_{\rm lab} = 200$~MeV. 
             \end{itemize}
\end{footnotesize}
\end{table}

Finally, consider the description of the scattering data in comparison 
to other available high-precision potentials.
Tab.~\ref{tab:chi2-emn-sms} gives the $\chi^2$/datum values of the
most recent chiral N$^4$LO$^+$ potentials, i.e. the SMS potential of
this work and the nonlocally regularized Entem-Machleidt-Nosyk
(EMN) potential of Ref.~\cite{Entem:2017gor}, for all available
cutoff values. When comparing the numbers, one should keep in mind
that the SMS interaction was explicitly fitted to this particular
database while the database employed for the EMN potentials slightly
differs from the one employed here, in particular with respect to the neutron-proton
data. The shown results nevertheless give a reliable
qualitative idea about the accuracy of the interactions.
\begin{table}[t]
    \centering
    \smallskip

    \begin{tabular*}{\textwidth}{@{\extracolsep{\fill}}lcccc}
        \toprule
        \noalign{}
        $E_\mathrm{lab}$ bin & CD Bonn & Nijm I & Nijm II & Reid93
%        \smallskip
        \\
        \midrule
        $0-100$ & 1.015 & 1.000 & 1.008 & 1.004 \\
        $0-200$ & 1.030 & 1.037 & 1.050 & 1.057 \\
        $0-300$ & 1.042 & 1.061 & 1.070 & 1.078 \\[0pt]
        \bottomrule
    \end{tabular*}
    \caption{$\chi^2$/datum for the description of the combined
     neutron-proton and proton-proton scattering data up to
     $E_\mathrm{lab} = 300$~MeV of the semi-phenomenological
     CD-Bonn \cite{Machleidt:2000ge} and Nijmegen NijmI, NijmII and
     Reid93 potentials \cite{Stoks:1994wp}.
    \label{tab:chi2-semi-phenomenological-potentials}}
\end{table}
This is supported by the results of Tab.~\ref{tab:chi2-semi-phenomenological-potentials}
which give the corresponding $\chi^2$/datum values for selected
high-precision semi-phenomenological potentials. Although the
Nijmegen potentials (NijmI, NijmII and Reid93) and the CD-Bonn
potential have been fitted to older 1993 and 1999 databases,
respectively, they hold up quite well when compared to the authors' own
data selection. It should be noted that when restricted to data before
1993/1999 the database used in Ref.~\cite{Entem:2017gor} should be
identical to these of the semi-phenomenological potentials. These
values also show that in the energy range $E_\mathrm{lab} =
0$--$300$~MeV, the SMS N$^4$LO$^+$ interaction achieves or even
exceeds the precision of the most sophisticated phenomenological
potentials.
%accuracy of previous high-precision potentials.

%***
For the sake of completeness, Tab.~\ref{tab:chi2-emn-sms} also provides a comparison of
the SMS and EMN chiral potentials at the N$^3$LO level. When comparing
the description of the data, one should keep in mind that the N$^3$LO SMS
chiral potentials have been fitted to the NN scattering data up to
$E_{\rm lab} = 200$~MeV only. Notice further that the EMN potentials 
employ partial-wave dependent functional form of the regulator for
contact interactions (and have $15$ order-$Q^4$ contact
interactions as compared to $12$ terms in the SMS interactions).

\subsection{\textit{Selected applications in the NN sector}}

\begin{figure}[tb]
  \vskip 1 true cm
  \begin{center}
    \includegraphics[width=\textwidth,keepaspectratio,angle=0,clip]{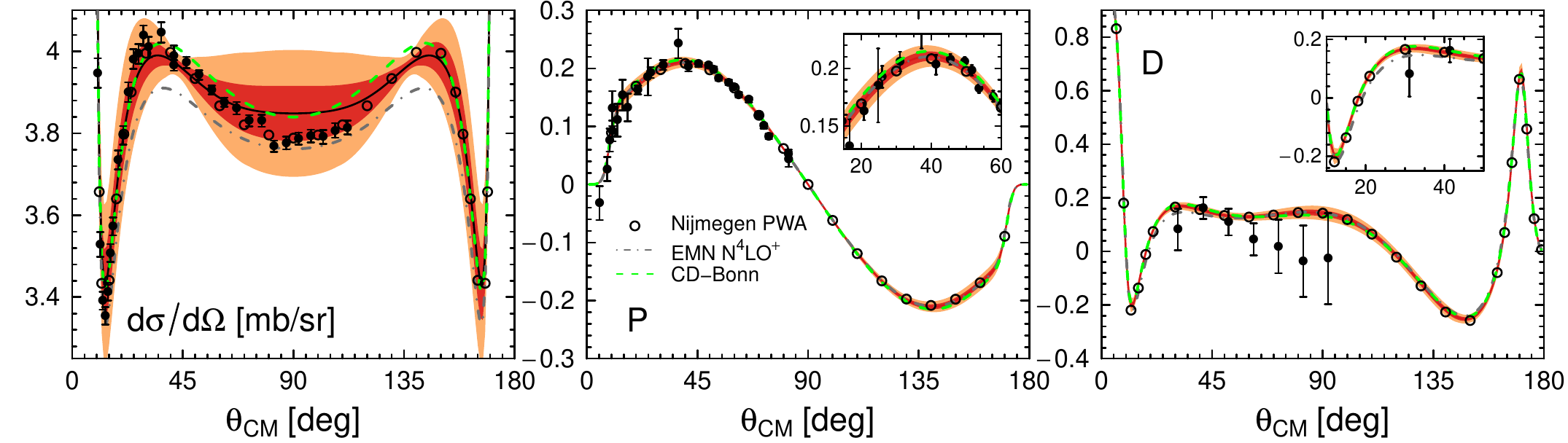}
  \end{center}
  \caption{Selected proton-proton observables around $E_\mathrm{lab} =
    143$~MeV. Differential cross section $d\sigma/d\Omega$ at
    $E_\mathrm{lab} = 144.1$~MeV are shown in the left panel with experimental data from
    Refs.~\cite{Cox:1968jxz,Jarvis:1971fla}. The data sets have been
    corrected for their estimated norms of 0.988 and 1.001,
    respectively, which have been taken from
    Ref.~\cite{Reinert:2017usi}. Analyzing power $P$ at
    $E_\mathrm{lab} = 142$~MeV are shown in the middle panel with experimental data of
    Ref.~\cite{Taylor:1960}, which has been corrected for its
    estimated norm of 0.940. The right panel shows depolarization $D$ at $E_\mathrm{lab} =
    143$~MeV with experimental data from Ref.~\cite{Bird:1961}. The
    dark and light red bands show the truncation error of the
    N$^4$LO$^+$ result with $\Lambda = 450$~MeV at the 68\%- and
    95\%-DoB, respectively, with the central value shown by a solid
    black line ($d\sigma/d\Omega$ only). Open black circles, gray
    dashed-dotted and green dashed lines denote the results of the
    Nijmegen PWA, the N$^4$LO$^+$ EMN potential with the central
    cutoff $\Lambda = 500$~MeV and the CD-Bonn potential,
    respectively. 
    \label{fig:pp-observables-143MeV}
  }
\end{figure}

As an example for the description of scattering data,  the SMS
N$^4$LO$^+$ results are shown for the cutoff $\Lambda = 450$~MeV along
with the
experimental data for selected proton-proton observables around
$E_\mathrm{lab} = 143$~MeV in Fig.~\ref{fig:pp-observables-143MeV}. 
The results of the Nijmegen PWA, the EMN potential with
the central cutoff $\Lambda = 500$~MeV and the CD-Bonn potential are
also provided for
the purpose of comparison. The differential cross section data of
Ref.~\cite{Cox:1968jxz} shown in the left panel has been used in
Ref.~\cite{Reinert:2017usi} to illustrate the importance of F-waves in
the description of high-precision proton-proton observables. The
differential cross section data of
Fig.~\ref{fig:pp-observables-143MeV} are not included in the authors' own
database selection, and the shown N$^4$LO$^+$ results are thus
predictions instead of being fitted. Nevertheless, the data are well
described within the truncation uncertainties at N$^4$LO$^+$. There
exists some variation in the predictions of the differential cross
section among the shown results by the other groups, especially when
compared with the good agreement for the two spin observables $P$ and
$D$. It should be noted that no definite conclusions regarding the
accuracy of the different results can be made based on the data in
Fig.~\ref{fig:pp-observables-143MeV} alone. The authors have chosen to show the
data with their estimated norms from Ref.~\cite{Reinert:2017usi},
which are in good agreement with the Nijmegen PWA. The norms of these
data sets estimated in the Granada 2013 PWA, on the other hand, would
bring the experimental data closer to the CD-Bonn result.

\begin{figure}[tb]
  \vskip 1 true cm
  \begin{center}
    \includegraphics[width=\textwidth,keepaspectratio,angle=0,clip]{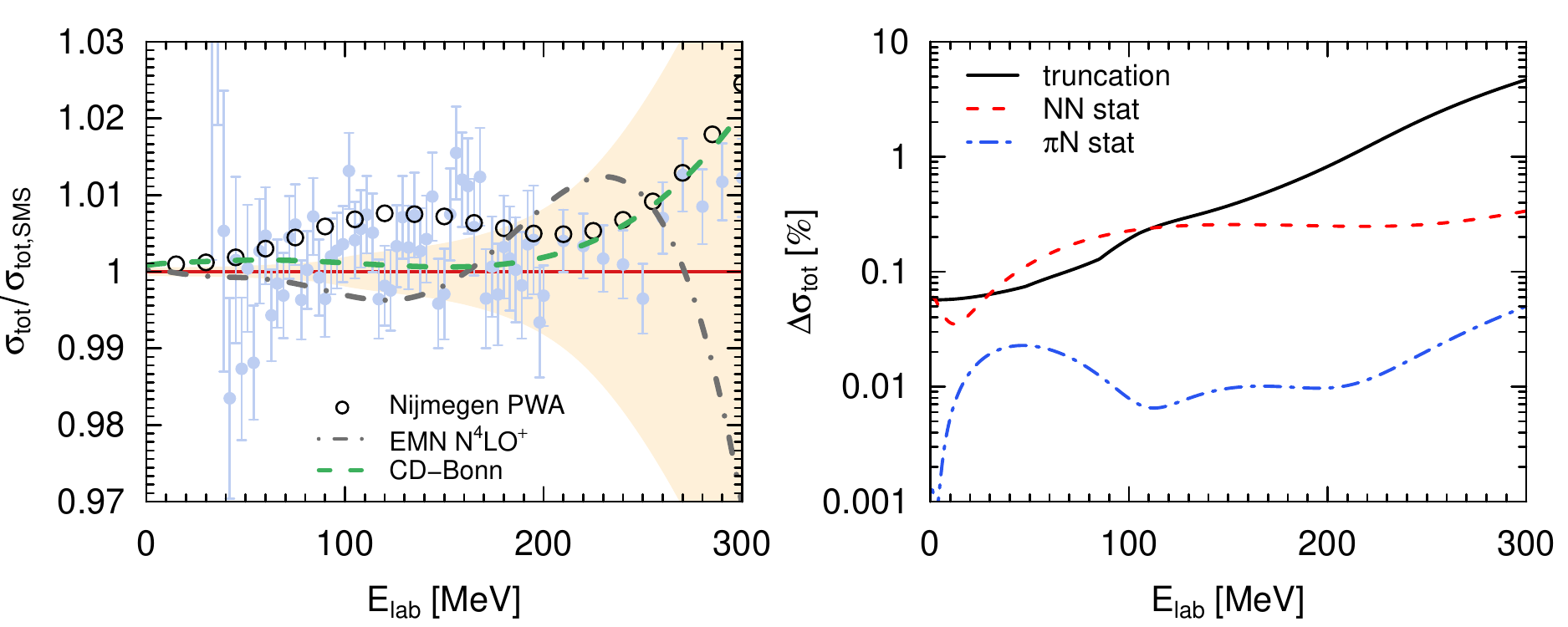}
  \end{center}
  \caption{Neutron-proton total cross section in the range
    $E_\mathrm{lab} = 0$--$300$~MeV. The left panel shows the total
    cross sections divided by the SMS N$^4$LO$^+$ results for $\Lambda =
    450$~MeV. The light blue points are the experimental data of
    Ref.~\cite{Lisowski:1982rm} corrected for an estimated norm of
    1.002. Peach-colored bands represent the truncation uncertainty
    for the 68\% DoB. For the remaining notation of this panel see
    Fig~\ref{fig:pp-observables-143MeV}. The right panel compares the
    different uncertainties of the SMS N$^4$LO$^+$ result for $\Lambda =
    450$~MeV: Black solid, red dashed and blue dashed-dotted lines
    denote the truncation uncertainty and the statistical
    uncertainties due to the parameters fitted from the NN system and
    the $\pi$N LECs of Ref.~\cite{Hoferichter:2015tha}, respectively. 
    \label{fig:total-cross-section}
  }
\end{figure}

Fig.~\ref{fig:total-cross-section} shows the neutron-proton total
cross sections in the energy range $E_\mathrm{lab} =
0$-$300$~MeV. Similar to Fig.~\ref{fig:pp-observables-143MeV}, the SMS
N$^4$LO$^+$ result for $\Lambda = 450$~MeV  is compared with other
predictions and selected experimental data in the left panel. The authors show
here relative values as the total cross section changes considerably
%in magnitude
in this energy range and the shown differences are small
compared to that scale. The right panel
%, on the other hand,
exemplifies the uncertainty quantification.
%performed for 
    %     results.
Here, the relative sizes of the truncation error, the
statistical error due to the parameters determined from NN data and
the statistical error due to the $\pi$N LECs of
Ref.~\cite{Hoferichter:2015tha} are shown. As expected based on the
phase shift differences in Fig.~\ref{fig:phase-shift-diff-np}, the NN
statistical error can become large at smaller energies below
$E_\mathrm{lab} = 100$~MeV while the truncation error dominates at
higher energies. The $\pi$N statistical error is about one order of
magnitude smaller than the NN statistical one.

\begin{table}[tb]
    \caption{Deuteron binding energy $B_d$, asymptotic S-state
      normalization $A_S$, asymptotic D/S-state ratio $\eta$, matter
      radius $r_m$, leading contribution to the quadrupole moment
      $Q_0$ and D-state probability $P_D$ for the SMS potential at
      N$^4$LO$^+$ for all values of the cutoff. The first error is
      the statistical uncertainty with respect to both NN and $\pi$N
      parameters and the second error is the truncation uncertainty
      (only provided for the observables $A_S$ and $\eta$). 
    \label{tab:deuteron-properties}}
    \smallskip
\begin{tabular*}{\textwidth}{@{\extracolsep{\fill}}lr@{\extracolsep{1pt}}l@{\extracolsep{\fill}}r@{\extracolsep{1pt}}l@{\extracolsep{\fill}}r@{\extracolsep{1pt}}l@{\extracolsep{\fill}}r@{\extracolsep{1pt}}l@{\extracolsep{\fill}}r@{\extracolsep{1pt}}l@{\extracolsep{1pt}}l}
        \toprule
                 & \multicolumn{2}{c}{$\Lambda = 400$ MeV} & \multicolumn{2}{c}{$\Lambda = 450$ MeV} & \multicolumn{2}{c}{$\Lambda = 500$ MeV} & \multicolumn{2}{c}{$\Lambda = 550$ MeV} & \multicolumn{3}{c}{Empirical} \\
        \midrule
        $B_d$ (MeV)                            & $2.2246$ &     $^\star$     & $2.2246$ &   $^\star$       & $2.2246$ &   $^\star$       & $2.2246$ &  $^\star$       & $2.22456614$ & $(41)$ & \cite{Kessler:1999zz} \\
        % $\langle T_\mathrm{kin} \rangle$ (MeV) &    $13.57$ &          &    $14.41$ &          &    $15.40$ &          &    $16.43$ &          & \\
        $A_S$ (fm$^{-1/2}$)                    &   $0.8842$ & $(3)(5)$ &   $0.8846$ & $(3)(5)$ &   $0.8848$ & $(3)(5)$ &   $0.8851$ & $(3)(6)$ &   $0.8845$ & $(8)$  & \cite{deSwart:1995ui} \\
        $\eta$                                 &   $0.0260$ & $(2)(1)$ &   $0.0261$ & $(2)(0)$ &   $0.0263$ & $(2)(0)$ &   $0.0265$ & $(2)(0)$ &   $0.0256$ & $(4)$  & \cite{Rodning:1990zz} \\
        $r_m$ (fm)                             &   $1.9647$ & $(7)$    &   $1.9662$ & $(6)$    &   $1.9674$ & $(6)$    &   $1.9686$ & $(6)$    &  --- &  & \\
        $Q_0$ (fm$^2$)                         &    $0.271$ & $(2)$    &   $0.275$  & $(2)$    &    $0.279$ & $(2)$    &    $0.282$ & $(2)$    & --- & & \\
        $P_D$ (\%)                             &     $4.25$ &          &     $4.79$ &          &     $5.29$ &          &     $5.73$ &          &        --- &        & \\
        \bottomrule
    \end{tabular*}
    \begin{tabular*}{\textwidth}{@{\extracolsep{\fill}}l}
        $^\star$The deuteron binding energy has been taken as input in the fit. %\\[-2pt]
  \end{tabular*}
\end{table}

The excellent reproduction of experimental data of the SMS N$^4$LO$^+$
interaction also extends to the deuteron bound state, whose properties
are shown in Tab.~\ref{tab:deuteron-properties} for all considered
cutoff values. The authors have limited the estimation of the truncation error
to (non-fitted) observables and thus provide them only for $A_S$ and
$\eta$. $r_m$ and $Q_0$ are related to the corresponding moments of the
probability density distribution and thus determined solely by the
deuteron wave function. These quantities are not measurable but 
constitute the leading contributions to the observable structure
radius and the quadrupole moment, which will be discussed below. 

\begin{figure}[tb]
  \begin{center}
\includegraphics[width=\textwidth,keepaspectratio,angle=0,clip]{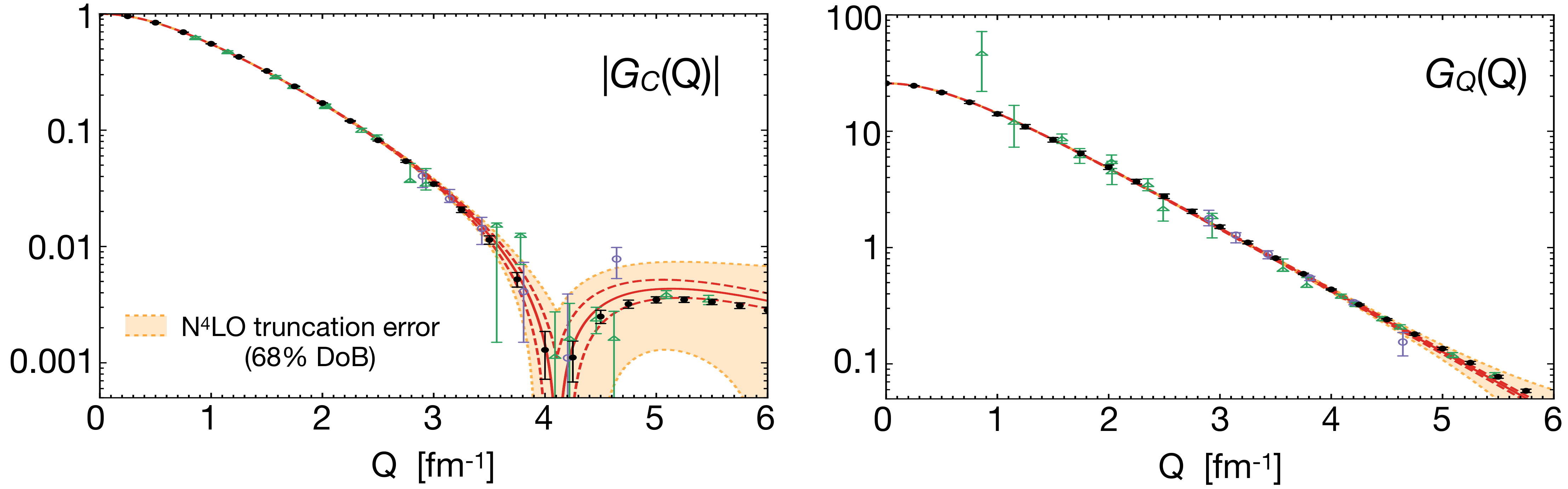}  
\end{center}
\caption{The deuteron charge (left panel) and quadrupole (right panel)
  form factors calculated at N$^4$LO for the cutoff $\Lambda =
  500$~MeV.  Bands between dashed red lines correspond to a $1\sigma$
  error in the determination of the 2N short-range contributions to
  the charge density operator. For references to the experimental data
  and their parametrization shown by the black solid circles see
  Ref.~\cite{Filin:2020tcs}. 
\label{fig_FF}
}
\end{figure}

It is interesting to test the novel SMS 2NF and more generally the
employed chiral EFT framework by calculating
the deuteron electromagnetic form factors (FF). The magnetic FF of the
deuteron requires the calculation of the expectation value of the
current density operator, whose consistently regularized 2N
contributions are only available at N$^2$LO. On the other hand, the charge
and quadrupole FFs $G_C (Q)$ and $G_Q (Q)$, with $Q^2 = - q^2 > 0$ and $q$
denoting the four-momentum of the virtual photon, 
depend on the isoscalar charge density
operator that has been worked out in chiral EFT to a high accuracy. 
In Fig.~\ref{fig_FF}, the results for the charge and quadrupole FF
of the deuteron are shown at  N$^4$LO
\cite{Filin:2019eoe,Filin:2020tcs}. Here, the 
empirical results were used for the nucleon form factors to parametrize the
single-nucleon contributions to the charge density operator without
relying on the chiral expansion. The
relativistic corrections and the contributions of the 2N charge
density operators were also taken into account. The latter depend on two LECs that have been
fixed from the experimental data for $G_C (Q)$ and $G_Q
(Q)$, see Fig.~\ref{fig_FF}. With all LECs being determined as described
above, a prediction for the deuteron structure
radius $r_{\rm str}$ and  the quadrupole moment $Q_d= G_Q
(0)$ was made. The structure radius denotes the contribution to the
deuteron charge radius, which is related to the FF $G_C$ via $r^2 = - 6 \frac{d G_C
  (Q^2)}{dQ^2}\Big|_{Q^2 = 0}$, that arises from the nuclear binding
mechanism. Up to the so-called Darwin term, $r_{\rm str}$ can be
interpreted as the charge radius of the deuteron made out of
structure-less nucleons. See Ref.~\cite{Filin:2020tcs} for more details and
Ref.~\cite{Epelbaum:2022fjc} for the interpretation of the FF in terms of the charge density
distribution.  

The final predictions for the deuteron structure radius and quadrupole
moment in Ref.~\cite{Filin:2020tcs} read
\beq
r_{\rm str} = 1.9729^{+0.0015}_{-0.0012}\; {\rm fm}, \quad \quad
Q_d = 0.2854^{+0.0038}_{-0.0017}\; {\rm fm}^2\,,
\eeq
where the quoted errors include the statistical uncertainties of various
LECs, the uncertainty in the parametrizations of the nucleon FFs
and the estimated N$^4$LO truncation uncertainty. These values are to be
compared with the determinations from laser spectroscopy
experiments \cite{Jentschura:2011,Puchalski:2020jkt}: 
\beq
r_{\rm str}^{\rm exp} = 1.97507(78)\; {\rm fm}, \quad \quad
Q_d^{\rm exp} = 0.285699 (23)\; {\rm fm}^2\,.
\eeq
Here, the quoted  value for $r_{\rm str}^{\rm exp} $ was obtained
using the mean square neutron radius of $r_n^2 = -0.114(3)$~fm$^2$.  
Alternatively, the prediction for $r_{\rm str}^{\rm exp}$ was used in
combination with the very precise experimental data on the
deuteron-proton charge-radius difference from
Ref.~\cite{Jentschura:2011} to update the value of the mean square 
neutron radius $r_n^2 = -0.105^{+0.005}_{-0.006}$~fm$^2$.

\section{\textit{Beyond the NN system}}

\subsection{\textit{SMS three-nucleon force at N$^2$LO}}

\begin{figure}[tb]
  \begin{center}
\includegraphics[width=\textwidth,keepaspectratio,angle=0,clip]{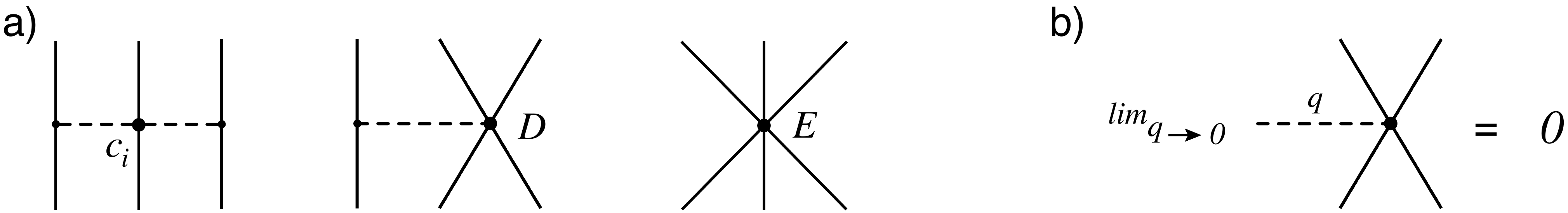}  
\end{center}
\caption{
        a) Feynman diagrams contributing to the 3NF at N$^2$LO. Solid
        and dashed lines refer to nucleons and pions, respectively.  b)
        Chiral symmetry constraint on the $\pi$NN interaction.
\label{fig_3NF}
}
\end{figure}

The leading 3NF at N$^2$LO is generated by the tree-level topologies
visualized in Fig.~\ref{fig_3NF} a), see also Tab.~\ref{tab:Forces}. 
It is straightforward to regularize the corresponding expressions
using the semi-local cutoff in momentum space and
employing the same convention for the subtraction terms as in the SMS 2NF
 of Ref.~\cite{Reinert:2017usi}:
\beqa
\label{leading}
V^{\rm 3N}_\Lambda &=& \frac{g_A^2}{8 F_\pi^4}\;  
\bigg\{
\frac{\vec \sigma_1 \cdot \vec q_1  \; \vec \sigma_3 \cdot \vec q_3}{(q_1^{\, 2} + M_\pi^2) \, (q_3^{\, 2} + M_\pi^2)}
%&\times &\Big[T_{13}  \big( - 4 c_1 M_\pi^2 
%+ 2 c_3 \, \vec q_1 \cdot \vec
%    q_3 \big)  
\Big[T_{13}  \big( 2 c_3 \, \vec q_1 \cdot \vec q_3 - 4 c_1 M_\pi^2 \big)  
+  c_4 T_{132} \, \vec q_1 \times \vec q_3 
  \cdot \vec \sigma_2  \Big] 
  % &&{} \mbox{\hskip 3.75 true cm}+
  \nn[4pt]
&&{} \hspace{1.0cm}+\;
C\, \frac{\vec \sigma_1 \cdot \vec q_1}{q_1^{\, 2} + M_\pi^2}
\Big( 2 c_3 \, T_{13}\, \vec \sigma_3 \cdot \vec q_1 + c_4 T_{132} \; \vec q_1 \times \vec \sigma_3 
  \cdot \vec \sigma_2  \Big)   \nn[4pt]
&&{} \hspace{1.0cm}+\;
C  \, \frac{\vec \sigma_3 \cdot \vec q_3}{q_3^{\, 2} + M_\pi^2}
\Big( 2 c_3 \, T_{13}\, \vec \sigma_1 \cdot \vec q_3
+ c_4 T_{132}\; \vec \sigma_1 \times \vec q_3 
  \cdot \vec \sigma_2  \Big)  \nn
%  &&{}\mbox{\hskip 3.75 true cm}+
&&{} \hspace{1.0cm}+\;
C^2 \, \Big( 2 c_3 \, T_{13} \, \vec \sigma_1 \cdot \vec \sigma_3 
+ c_4 T_{132} \; \vec \sigma_1 \times \vec \sigma_3 
\cdot \vec \sigma_2 \Big)  \bigg\}
\;  e^{- \frac{q_1^2 +
M_\pi^2}{\Lambda^2}}\, e^{- \frac{q_3^2 + M_\pi^2}{\Lambda^2}}
\nn[4pt]
&-& \frac{g_A \, D}{8 F_\pi^2}\; T_{13} \;
 \bigg[
\frac{\vec \sigma_3 \cdot \vec q_3 }{q_3^{\, 2} + M_\pi^2} \; 
\vec \sigma_1 \cdot \vec q_3  
+ 
C  \, \vec \sigma_1 \cdot \vec \sigma_3 \bigg]\;
e^{- \frac{p_{12}^2 +p_{12}^{\, \prime 2}
}{\Lambda^2}}\,
  e^{- \frac{q_3^2 +
M_\pi^2}{\Lambda^2}}\, 
\nn[4pt]
&+&  
\frac{1}{2} E \; T_{12} \, e^{- \frac{p_{12}^2 +p_{12}^{\, \prime 2}
  }{\Lambda^2}}\, e^{- \frac{3 Q_{3}^2 +3 Q_{3}^{\, \prime 2}
}{4\Lambda^2}} 
\; + \;  
\mbox{5 permutations}\,, 
\eeqa
where $\vec q_{i} = \vec p_i \, ' - \vec p_i$ is the momentum transfer
of  nucleon $i$ with $\vec p_i $ and $\vec p_i \, '$ being the corresponding
initial and final momenta, respectively. Further, 
$\vec p_{12} = (\vec p_1 - \vec p_2)/2$ and $\vec Q_3 = 2 (\vec p_3 - (\vec p_1 + \vec p_2 )/2)/3$ are the Jacobi momenta  in the initial state and 
$\vec p_{12}^{\, \prime} = (\vec p_1^{\, \prime} - \vec p_2^{\, \prime})/2$ and $\vec Q_3^{\, \prime} = 2 (\vec p_3^{\, \prime} - (\vec p_1^{\, \prime} + \vec p_2^{\, \prime} )/2)/3$
in the final state. The isospin operators $T_{ij}$ and $T_{ijk}$
are defined as $T_{ij} \equiv \bm \tau_i \cdot \bm \tau_j$ and $T_{ijk} \equiv \bm
\tau_i \times \bm \tau_j \cdot \bm \tau_k$.
%Finally, $g_A$, $F_\pi$ and $M_\pi$ refer to the nucleon axial vector
    %     coupling, pion decay constant and pion mass, respectively, while
The
subtraction constant $C$ is specified in Eq.~(\ref{subtr}).
%The
%subtraction terms proportional to the constant  $C$,
%\beq
%C = -\frac{\Lambda \left(\Lambda ^2-2 M_\pi^2 \right) + 2 \sqrt{\pi } M_\pi^3 e^{\frac{M_\pi^2}{\Lambda ^2}}
%  {\rm erfc}\left(\frac{M_\pi}{\Lambda }\right)}{3 \Lambda ^3} \,,
% \eeq
%ensure that the corresponding $r$-space potentials are regular at the
%origin. Here, ${\rm erfc} (x)$ denotes the complementary error function. 

\subsection{\textit{Nucleon-deuteron scattering}}

Three-nucleon scattering and bound state observables are calculated
by solving the Faddeev equations in momentum space in the partial wave
basis, see Ref.~\cite{Maris:2020qne} for details.  The partial wave
decomposition of a general 3NF is carried out numerically as detailed
in Ref.~\cite{Hebeler:2015wxa}. 

\begin{figure}[tb]
  \begin{center}
\includegraphics[width=\textwidth,keepaspectratio,angle=0,clip]{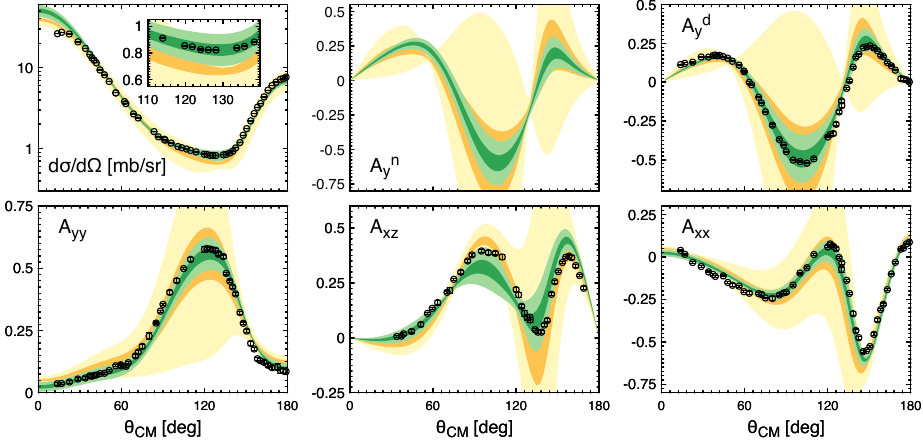}  
\end{center}
\caption{
\label{fig_Nd} Results for selected observables in elastic
  Nd scattering  at laboratory energy of
  $E_N = 70$~MeV  at NLO
  (yellow bands) and N$^2$LO (green bands) for $\Lambda =
  500$~MeV. 
The light- (dark-) shaded bands indicate the estimated $95\%$ ($68\%$)
DoB intervals.
%using the Bayesian model $\bar C_{0.5-10}^{650}$. 
  Open circles are proton-deuteron data from
  Ref.~\cite{Sekiguchi:2002sf}. 
}
\end{figure}

The leading 3NF depends on the LECs $D$ and $E$, whose values are
extracted from low-energy 3N observables. Specifically, following
Refs.~\cite{LENPIC:2018ewt,Maris:2020qne}, the authors first require that the $^3$H binding energy is
correctly reproduced. This constraint fixes the value of the LEC $E$
for a given value of $D$. To reliably determine this remaining LEC, it is
essential to employ observables that are not strongly correlated 
with the $^3$H binding energy \cite{Wesolowski:2021cni}. The authors
of Ref.~\cite{LENPIC:2018ewt} have
studied the constraints imposed on the value of the LEC $D$ by the experimental data for the Nd
doublet scattering length, the total Nd scattering cross section and
the differential cross section minimum in elastic Nd scattering at
several energies.
It was found that the high-precision experimental data of Ref.~\cite{Sekiguchi:2002sf} for
the differential cross section at the proton energy of $E_p = 70$~MeV,
see Fig.~\ref{fig_Nd}, 
provide a particularly strong constraint on $D$. This finding is
in line with the known sensitivity of the cross section minimum to 3NF
effects, see \cite{Gloeckle:1995jg} and references therein. 
The LECs $D$ and $E$ determined from the $^3$H binding energy and the 
Nd cross section minimum at $70$~MeV were shown to result in a consistent
description of all observables mentioned above, see Fig.~2 of
\cite{LENPIC:2018ewt} and Fig.~2 of \cite{Maris:2020qne}.  

Having determined the LECs $D$ and $E$ as described above, it is
interesting to consider selected predictions for Nd scattering observables up
to N$^2$LO. In Fig.~\ref{fig_Nd}, it is demonstrated that the NLO and
N$^2$LO results for the differential cross section and selected polarization
observables in elastic Nd scattering at $70$~MeV are in agreement with
the experimental data. For the  analyzing powers
$A_{yy}$ and $A_{xz}$,  the discrepancies between the calculations and
the experimental data are comparable with the truncation
errors at  N$^2$LO. These discrepancies are not resolved by
higher-order corrections to the 2NF \cite{Epelbaum:2019zqc} and thus indicate the
important role played by subleading 3NF contributions. The residual
cutoff dependence of the calculated 3NF is, in general, compatible
with the estimated truncation errors. In particular, the results for
$\Lambda = 500$~MeV shown in Fig.~\ref{fig_Nd} are very similar to
those using  $\Lambda = 450$~MeV and shown in Fig.~3 of
Ref.~\cite{Maris:2020qne}. The dependence of the obtained predictions on the
different choices of semi-local regulators (i.e., coordinate-space versus
momentum-space) and the subtraction conventions for the 3NF is also
fully consistent with the truncation uncertainty, see
\cite{Maris:2020qne,LENPIC:2018ewt} for details. The estimated
truncation errors for Nd scattering observables were further validated
by analyzing the contributions of selected higher-order short-range
3NF terms in Ref.~\cite{Epelbaum:2019zqc}.   

\begin{figure}[tb]
  \begin{center}
\includegraphics[width=\textwidth,keepaspectratio,angle=0,clip]{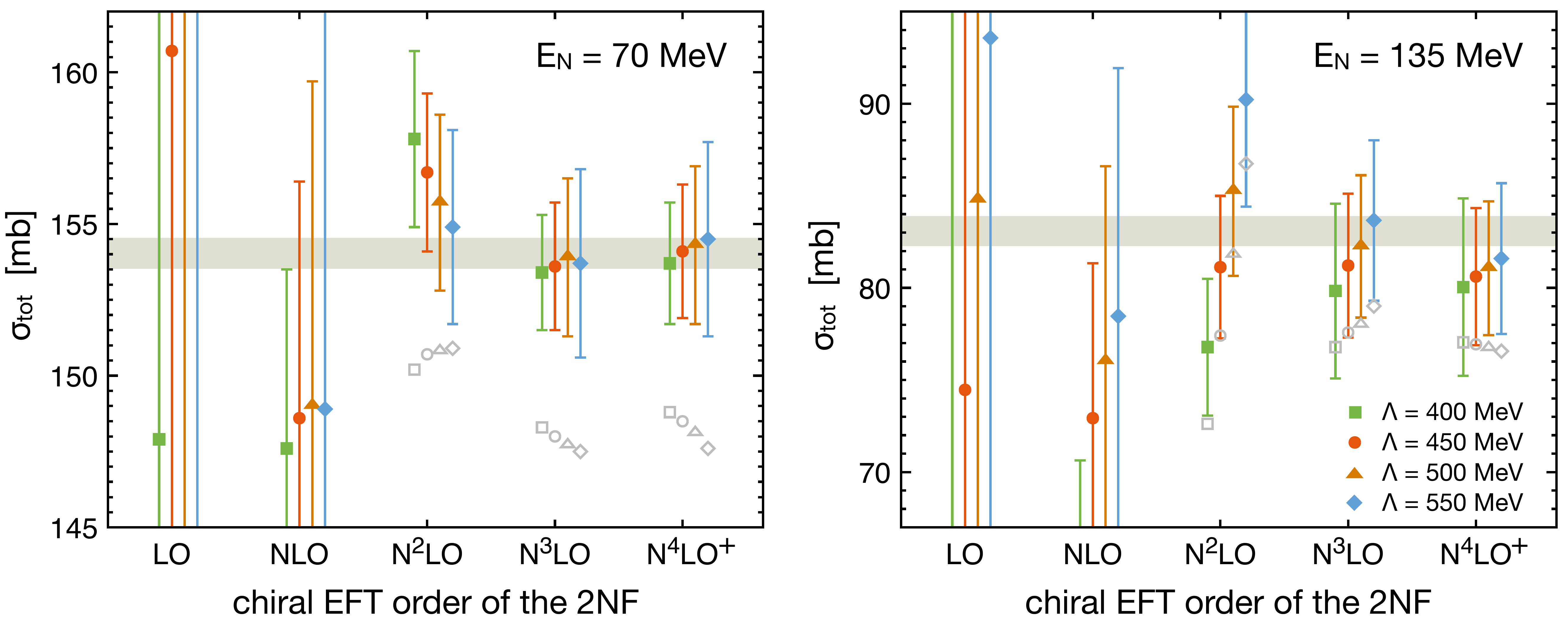}  
\end{center}
\caption{Predictions for the Nd total cross section at $70$~MeV (left
  panel) and $135$~MeV (right panel) based on the SMS chiral
  interactions at different orders (shown by solid symbols with error
  bars). 3NF is included at N$^2$LO only. Error bars show the EFT
  truncation uncertainty
  %estimated using the $\bar
  % C_{0.5-1.0}^{650}$ Bayesian model from Ref.~\cite{Epelbaum:2019zqc}
  (68\% DoB
  intervals). For the incomplete calculations at N$^3$LO and N$^4$LO$^+$, the
quoted errors are the N$^2$LO truncation uncertainties. Gray open symbols without
  error bars show the results based on the 2NF only.  Horizontal bands
  are experimental data from Ref.~\cite{Abfalterer:1998zz}. 
\label{fig_Nd_cs}
}
\end{figure}

Finally, in Fig.~\ref{fig_Nd_cs}, the predictions for the total cross
section at $E_N = 70$ and $135$~MeV are shown for all four cutoff values. These
calculations are based on the SMS 2NF at LO, NLO, N$^2$LO, N$^3$LO and
N$^4$LO$^+$ from Ref.~\cite{Reinert:2017usi}. Starting from N$^2$LO, the leading 3NF contributions specified in
Eq.~(\ref{leading}) are also taken into account.  For each combination of the 2NF and 3NF and for
each cutoff value, the LECs $D$ and $E$ are fixed to reproduce the
$^3$H binding energy and the Nd cross section minimum at $70$~MeV as
described above. Since the 3NF is included only at N$^2$LO, one can regard
the predictions based on the 2NF at  N$^3$LO and N$^4$LO$^+$ as
alternative N$^2$LO calculations when estimating the truncation
error. The total cross section is underestimated by $\sim 3.5\%$ ($\sim
7\%$) at $E_N = 70$~MeV ($E_N = 135$~MeV) when 
using the high-precision 2NF at N$^4$LO$^+$ alone. Similar discrepancies
that tend to increase with the energy were also observed in calculations
based on high-precision phenomenological 2N potentials
\cite{Abfalterer:1998zz}. Adding the leading 3NF is
essential to bring the chiral EFT predictions at N$^2$LO in agreement
with the data. Moreover, the 3NF contributions to the total cross
section appear to be comparable to the NLO truncation errors in
agreement with the Weinberg power counting \cite{Weinberg:1991um}. 
Also the differences between the N$^2$LO and N$^3$LO
predictions that originate from the N$^3$LO contributions to the 2NF are
comparable with the estimated N$^2$LO truncation errors.

\subsection{\textit{Heavier systems}}

\begin{figure}[tb]
  \begin{center}
\includegraphics[width=\textwidth,keepaspectratio,angle=0,clip]{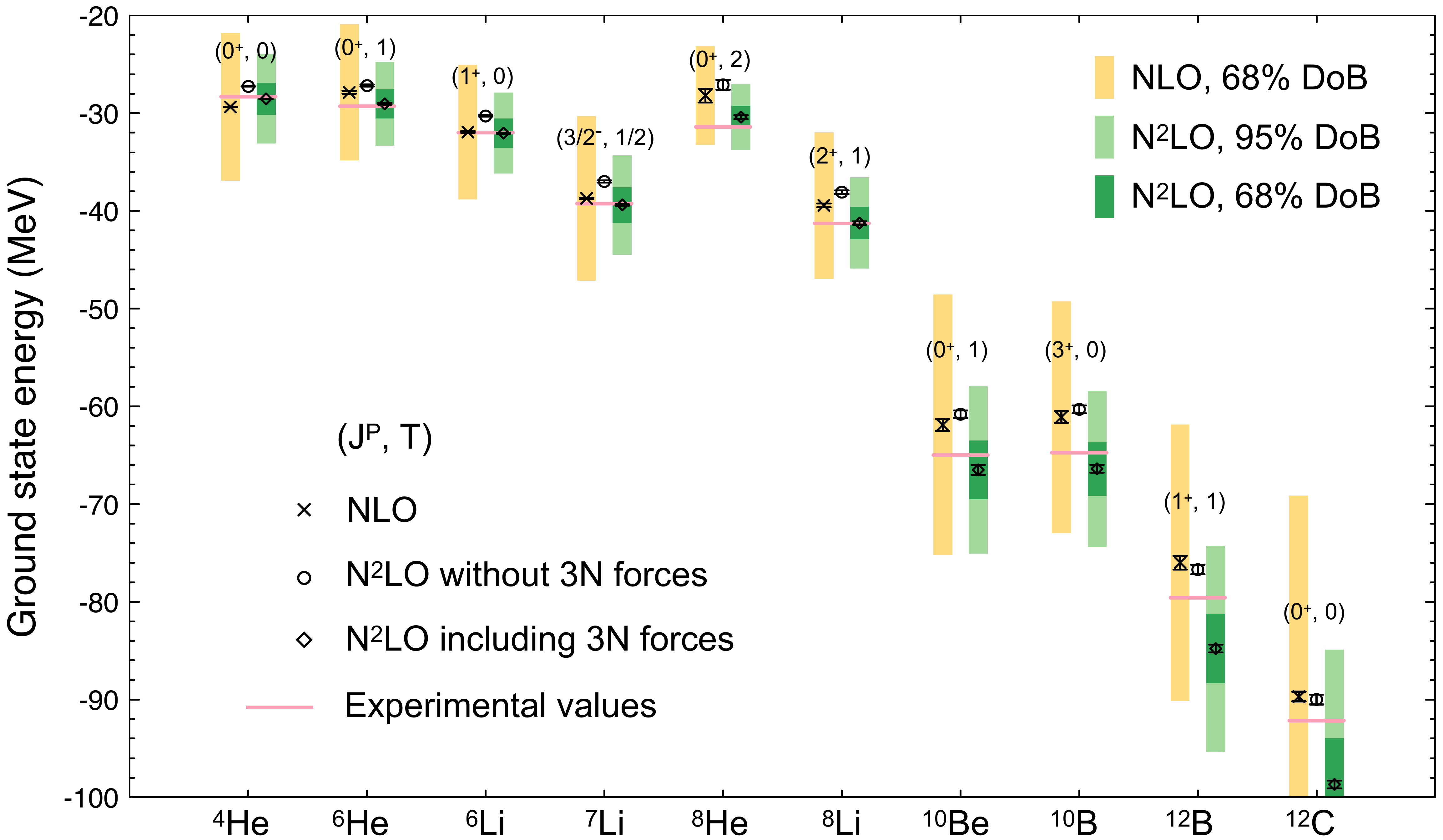}  
\end{center}
\caption{Predictions for ground state energies of selected nuclei with
  $A=4\, $--$
  12$ at NLO and N$^2$LO for $\Lambda = 450$~MeV using the
  ab-initio No-Core Configuration Interaction method (NCCI). Black error bars
  indicate the NCCI uncertainties, while shaded bars refer to the EFT
  truncation errors (not shown for incomplete N$^2$LO calculations
  based on 2NF only). See Ref.~\cite{Maris:2020qne} for details. 
\label{fig_spectra}
}
\end{figure}

The SMS chiral interactions have also been applied by the LENPIC collaboration
to predict 
the ground and excited state energies and radii of light and
medium-mass nuclei.  In Fig.~\ref{fig_spectra}, 
the results for the ground state energies of selected
nuclei up to $A=12$ are shown using the NLO 2NF (left symbols), the N$^2$LO
2NF (middle symbols) and the N$^2$LO 2NF in combination with the
N$^2$LO 3NF (right symbols) for  the cutoff $\Lambda = 450$~MeV as a
representative example.
Notice that the calculated energies are pure predictions since the LECs in the nuclear
Hamiltonian are determined from the 2N and 3N systems only.
For nuclei with $A \le 10$, the leading 3NF significantly improves the
agreement with the data by increasing the binding energy. For heavier
nuclei, the N$^2$LO Hamiltonian is systematically too attractive (but
the ground state energies of both $^{12}$B and $^{12}$C are still in
agreement with the data within $1.5 \sigma$). This tendency continues
and increases with the mass number \cite{Maris:2020qne}. The origin of
the systematic overbinding in heavier nuclei is currently under
investigation by the LENPIC collaboration. More results for $p$-shell
nuclei based on semi-local chiral EFT interactions can be found in
Refs.~\cite{LENPIC:2015qsz,LENPIC:2018lzt,LENPIC:2018ewt,Maris:2020qne}.

\section{\textit{Towards consistent regularization beyond the 2N system}}

The missing three- and four-nucleon forces and exchange current
operators at N$^3$LO and beyond are becoming more and more of a
bottleneck for {\it ab initio} low-energy nuclear theory. Although 
the N$^3$LO and even some of the N$^4$LO contributions to the many-body
forces and currents have been worked out using dimensional
regularization (DR), see Tab.~\ref{tab:Forces},  the existing expressions can {\it not} be
directly employed in few-nucleon calculations due to the
inconsistencies caused by combining the dimensional and cutoff
regularizations \cite{Epelbaum:2019kcf}. Below, an explicit example
will be given to
demonstrate such an inconsistency for the 3NF regularized in a naive way
using both (semi-) local and nonlocal cutoffs.

\subsection{\textit{Statement of the problem}}

Both the 2NF and 3NF need to be regularized in order to obtain a well defined solution of
the Faddeev equations.
%Due to underlying symmetries of the chiral
%Lagrangian, like e.g. chiral symmetry, the regulator has to respect
    %     these symmetries.
High-momentum components in the integrals appearing in the iterations of the Faddeev equation
generate contributions involving positive powers and logarithms of the
cutoff which diverge in the $\Lambda \to \infty$ limit and
are supposed to get absorbed by the available short-range
interactions. The momentum dependence of such contact interactions beyond the 2N sector is,
however, severely constrained by the spontaneously broken chiral symmetry of QCD. In
particular, in the limit of exact chiral symmetry (i.e., for $M_\pi
\to 0$), only derivative pion couplings are allowed in the effective
Lagrangian according to the Goldstone theorem.
In the 2N sector, the tree-level short range
interactions do not involve any pion couplings, and their momentum
dependence is therefore not restricted by the chiral symmetry. This is
in contrast to the $D$-like 3NF interactions, which are constrained by
the chiral symmetry as visualized in Fig.~\ref{fig_3NF}~b). These
constraints lead to inconsistencies that plague 
calculations involving the 3NF derived using DR and regularized
additionally by multiplying with
a local or nonlocal cutoff.
%to make the Faddeev equation well defined. 
As will be exemplified below, the resulting mismatch  between the two regularization
schemes can {\it not} be compensated by shifting the values of the available LECs.

To be specific, the authors focus here on the example discussed in
Ref.~\cite{Epelbaum:2019kcf} and consider the 
relativistic correction to $2\pi$-exchange 3NF proportional to
$g_A^2$~\cite{Bernard:2011zr}:
\beqa
V_{\rm 3N}^{2\pi, \, 1/m} &=& i\frac{g_A^2}{32 m
  F_\pi^4}\frac{\vec{\sigma}_1\cdot\vec{q}_1\,\vec{\sigma}_3\cdot\vec{q}_3}{(q_1^2+M_\pi^2)(q_3^2+M_\pi^2)}\bm{\tau}_1\cdot(\bm{\tau}_2\times\bm{\tau}_3)(2\vec{k}_1\cdot\vec{q}_3+4\vec{k}_3\cdot\vec{q}_3\nonumber\\
&+&i\,[\vec{q}_1\times\vec{q}_3]\cdot\vec{\sigma}_2) 
\; + \;  
\mbox{5 permutations}\,,\quad\quad\label{oneovermV2pigA2}
\eeqa
with $\vec{k}_i= \big(\vec{p}_i^{\,\prime}+\vec{p}_i\big)/2$.
Consider now the first iteration of these N$^3$LO contributions
with the LO $1\pi$-exchange 2N potential
\beq
 V_{\rm 2N}^{1\pi }= - \left(\frac{g_A}{2 F_\pi}\right)^2\bm{\tau}_1\cdot\bm{\tau}_2\frac{\vec{\sigma}_1\cdot\vec{q}\,\vec{\sigma}_2\cdot\vec{q}}{q^2 + M_\pi^2}.
\eeq
Regularization of these 2NF and 3NF is achieved by multiplying them with a
local Gaussian cutoff 
\beq
V_{\rm 3N, \, \Lambda}^{2\pi, \, 1/m} \, = \, V_{\rm 3N}^{2\pi,
  \, 1/m} \; e^{-\frac{q_1^2+M_\pi^2}{\Lambda^2}}\, 
e^{-\frac{q_3^2+M_\pi^2}{\Lambda^2}},
\quad \quad
V_{\rm 2N, \, \Lambda}^{1\pi }\,=\, V_{\rm 2N}^{1\pi } \; e^{-\frac{q^2+M_\pi^2}{\Lambda^2}} \,.
\eeq
Performing  a
large-$\Lambda$ expansion leads to  
\beqa
V_{\rm 3N, \, \Lambda}^{2\pi, \, 1/m} \, G_0  \, V_{\rm 2N, \,
  \Lambda}^{1\pi } &+& V_{\rm 2N, \,
  \Lambda}^{1\pi }  \, G_0  \, V_{\rm 3N, \, \Lambda}^{2\pi, \, 1/m}  
 \nn
&=&\Lambda\frac{g_A^4}{128\sqrt{2}\pi^{3/2}F_\pi^6}(\bm \tau_2\cdot \bm
\tau_3-\bm \tau_1\cdot \bm \tau_3)
\frac{{\vec{q}_2}\cdot\vec{\sigma}_2
  \vec{q}_3\cdot\vec{\sigma}_3}{q_3^2+M_\pi^2} \nn
&-&\Lambda\frac{g_A^4}{96\sqrt{2}\pi^{3/2}F_\pi^6} \, \bm{\tau}_1\cdot
  \bm{\tau}_3\frac{\vec{q}_3\cdot\vec{\sigma}_3
  \vec{q}_3\cdot\vec{\sigma}_1}{q_3^2+M_\pi^2}+\dots\;,\label{chiralsymmetryviolatingDivergence} 
\eeqa
where $G_0$ is the free resolvent operator and the ellipses refer to all permutations of the nucleon labels and
to terms that are finite in the $\Lambda \to \infty$-limit. The last term on the
right-hand side (rhs) of
Eq.~(\ref{chiralsymmetryviolatingDivergence}) has the form of the
$D$-term of the N$^2$LO 3NF, and it therefore can be absorbed into a
redefinition of the LEC $D$. In contrast, the first term on the rhs of
Eq.~(\ref{chiralsymmetryviolatingDivergence}) can not be
absorbed into a redefinition of the LECs entering the N$^2$LO 3NF
since the corresponding structure is not allowed by the chiral
symmetry. One therefore expects this problematic term to cancel against
some other contribution in the 3N amplitude. The only other term 
%contribution
with the desired combination of the LECs is the 
$1\pi$-$2\pi$-exchange 3NF at N$^3$LO $\propto g_A^4$, whose
expression was derived in  Ref.~\cite{Bernard:2007sp}
using DR. Had one used the same cutoff regularization also in the
calculation of this 3NF, one would obtain a linearly divergent term 
\beqa
V _{\rm 3N,  \, \Lambda}^{2\pi-1\pi} &=& -\Lambda\frac{g_A^4}{128\sqrt{2}\pi^{3/2}F_\pi^6}(\bm
\tau_2\cdot \bm \tau_3-\bm \tau_1\cdot \bm \tau_3)
\frac{{\vec{q}_2}\cdot\vec{\sigma}_2
  \vec{q}_3\cdot\vec{\sigma}_3}{q_3^2+M_\pi^2}\nn
&-&\Lambda\frac{g_A^4}{32\sqrt{2}\pi^{3/2}F_\pi^6} \, \bm{\tau}_1\cdot
\bm{\tau}_3 \frac{\vec{q}_3\cdot\vec{\sigma}_3
  \vec{q}_3\cdot\vec{\sigma}_1 }{q_3^2+M_\pi^2}+\dots\;,\label{twopiononepionDivergence}
%+\dots\;,\label{twopiononepionDivergence}
\eeqa
where the ellipses refer to terms which are finite in the $\Lambda \to
\infty$-limit that coincides with the DR result of
Ref.~\cite{Bernard:2007sp}. As expected, the problematic term in 
Eq.~(\ref{chiralsymmetryviolatingDivergence}) cancels exactly by the first
term on the rhs of Eq.~(\ref{twopiononepionDivergence}), while the
second term can, again, be absorbed into a redefinition of the LEC
$D$. Clearly, the cancellation is only operative if one uses the same
cutoff regularization in the derivation of the N$^3$LO 3NF and in
the Faddeev equation. A naive approach by multiplying the 
N$^3$LO 3NF expressions, derived using DR, with some cutoff regulators
obviously fails to ensure the cancellation of the chiral-symmetry-violating UV divergences
and results in an uncontrolled approximation for the amplitude, which 
cannot be renormalized. 

It is important to emphasize that the above inconsistency is
by no means restricted to the usage of local regulators for long-range interactions. 
Indeed, repeating the same exercise using a nonlocal regulator of a
Gaussian type,
\beq
V_{\rm 3N, \, \Lambda}^{2\pi, \, 1/m} \, = \, V_{\rm 3N}^{2\pi,
  \, 1/m} \; e^{-\frac{P_{12}^2+P_{12}^{\prime\,2}}{\Lambda^2}}\, 
e^{-3\frac{Q_{3}^2+Q_{3}^{\prime\,2}}{4\Lambda^2}},
\quad \quad
V_{\rm 2N, \, \Lambda}^{1\pi }\,=\, V_{\rm 2N}^{1\pi } \;
e^{-\frac{P_{12}^2+P_{12}^{\prime\,2}}{\Lambda^2}} \,,
\eeq
one obtains for the first iteration of the Faddeev equation with the
$1\pi$-exchange 2NF being antisymmetrized in the $12$-subsystem:
\beqa
V_{\rm 3N, \, \Lambda}^{2\pi, \, 1/m} \, G_0  \, V_{\rm 2N, \,
  \Lambda}^{1\pi } &+& V_{\rm 2N, \,
  \Lambda}^{1\pi }  \, G_0  \, V_{\rm 3N, \, \Lambda}^{2\pi, \, 1/m}  
  \nn
&=&\Lambda\frac{g_A^4}{1536\,(2\pi)^{3/2}F_\pi^6}\, i\,\bm \tau_1\cdot (\bm
\tau_2\times\bm\tau_3)
\frac{(7\,\vec{k}_1-3\,\vec{k}_2)\cdot\vec{\sigma}_1
  \vec{q}_3\cdot\vec{\sigma}_3}{q_3^2+M_\pi^2} \nn
&-&\Lambda\frac{g_A^4}{384\,(2\pi)^{3/2}F_\pi^6}\, \bm{\tau}_1\cdot
\bm{\tau}_3
\frac{\vec{q}_3\cdot\vec{\sigma}_3
  \vec{q}_3\cdot\vec{\sigma}_1 }{q_3^2+M_\pi^2}+\dots\;.\label{chiralsymmetryviolatingDivergenceNonlocal} 
\eeqa
Again, the first term on the rhs of
Eq.~(\ref{chiralsymmetryviolatingDivergenceNonlocal})  violates the
chiral symmetry and can not be absorbed into a redefinition of the LEC
$D$. The non-locality of the regulator thus does not cure the problem. 
It actually introduces additional complications by affecting the analytic
structure of the long-range potentials and making the derivation of
consistent cutoff-regularized 3NF technically more demanding.

The issue with inconsistent regularization affects not only three- and
more-nucleon forces, but it is also relevant for exchange currents at
and beyond N$^3$LO. In particular, analogous considerations for 
the axial vector current at N$^3$LO demonstrate the appearance of
chiral-symmetry-violating UV divergences when mixing the DR and cutoff
regularization \cite{Krebs:2020pii}.

\subsection{\textit{Possible solutions}}
The above inconsistencies can be rectified by using the
same regulator in the derivation of the nuclear forces and currents and
iterations of the dynamical equation. Such a regulator has to respect
all the
relevant symmetries. One option is to implement the regulator at the
level of the effective Lagrangian. One can, in particular, require
for the regularized pion propagator to take the form
$\exp (-\frac{q^2+M_\pi^2}{\Lambda^2} )/(q^2+M_\pi^2)$. 
This can be achieved by adding specific higher-derivative terms to the
effective Lagrangian, which disappear in the limit 
$\Lambda \to \infty$. Such higher-derivative regularization was
introduced by Slavnov a long time ago to regularize the
nonlinear sigma model~\cite{Slavnov:1971aw}. This idea can be used 
to construct a $\Lambda$-dependent effective Lagrangian for pions that
is manifestly invariant under global chiral transformations and yields 
long-range nuclear interactions regularized in a local way. On top of it,
one can employ a  nonlocal higher-derivative regularization for
contact interactions. The implementation of these ideas
in the 2N and 3N sectors is in progress.

Another possibility to implement  a symmetry preserving regulator
is given by the so-called gradient flow regularization approach
proposed originally by L\"uscher~\cite{Luscher:2010iy}, see
also~\cite{Grabowska:2015qpk,Kaplan:2016dgz}. The idea behind this
method is similar to that of the stochastic quantization by introducing a fifth dimension. The
original pion field that depends on space-time coordinates is to be
replaced by a field that depends, in addition,
on a fictitious time $t$. This field satisfies a gradient flow equation and
reduces to the original pion field in the limit $t\to 0$. Chiral
perturbation theory with this kind of regulator was discussed 
in Ref.~\cite{Bar:2013ora}, see also a related work in Ref.~\cite{Aoki:2014dxa}.

Last but not least, one can also employ a lattice regularization in 
chiral EFT. The regularized version of the effective pion Lagrangian on
the lattice can be found in~\cite{Borasoy:2003pg}. Nuclear forces and currents
regularized in this way are guaranteed to respect the underlying
symmetries and may be particularly useful for ongoing efforts to
extend nuclear lattice EFT simulations to higher orders \cite{Lahde:2019npb,Lee:2020meg}. 
%\begin{itemize}
%\item[--]  briefly explain stochastic regularization and the
%  higher-derivative regularization.   
%\end{itemize}

%{\color{red} HK, 1-2 pages.}

\section{\textit{Summary and outlook}}

To summarize, this chapter focused on the new generation of nuclear
interactions derived from chiral EFT using semi-local regulators
\cite{Epelbaum:2014efa,Epelbaum:2014sza,Reinert:2017usi,Reinert:2020mcu}.
At N$^4$LO$^+ $, the highest order 
available, the resulting SMS potentials were used to perform, for the
first time, a full-fledged partial wave analysis of proton-proton and
neutron-proton scattering data in the framework of chiral EFT \cite{Reinert:2020mcu,Reinert:2022thesis}. The
resulting near perfect description of NN data up to  the pion production
threshold leaves little room for improvement and suggests no need to
extend the EFT expansion beyond N$^4$LO$^+$ given the available NN data. 
The recent high accuracy calculation of the deuteron
charge and quadrupole form factors \cite{Filin:2019eoe,Filin:2020tcs}
were also briefly reviewed. 

The novel SMS interactions have been used to analyze 3N
scattering observables and selected properties of light and
medium-mass nuclei
\cite{Epelbaum:2019zqc,LENPIC:2015qsz,LENPIC:2018lzt,LENPIC:2018ewt,Maris:2020qne}.
The accuracy of these studies is limited by that of
the 3NF, which is only available at N$^2$LO. At this chiral order, the
predicted Nd scattering observables and ground state energies of 
nuclei with $A \le 12$ agree with the data within truncation error. 

N$^3$LO and some of the N$^4$LO contributions to  the 3NF and 4NF have
already been derived using dimensional regularization
\cite{Epelbaum:2005bjv,Epelbaum:2007us,Bernard:2007sp,Bernard:2011zr,Girlanda:2011fh,Krebs:2012yv,Krebs:2013kha},
see Tab.~\ref{tab:Forces}. Unfortunately,
these expressions cannot be employed to calculate
observables. This is because mixing the dimensional and cutoff regularizations
when calculating scattering amplitudes violates the chiral symmetry
and results in uncontrolled approximations beyond the 2N system.
This issue is not restricted to a particular type of cutoff regulator and applies to
local, semi-local and nonlocal cutoffs. It also plagues calculations
involving exchange currents at N$^3$LO and beyond. A 
solution of this challenge requires a complete
re-derivation of many-body forces and exchange currents using a
cutoff regulator that respects the underlying symmetries
such as e.g.~the higher-derivative regularization \cite{Slavnov:1971aw}. Work along this
line is in progress and will open an avenue for performing high-accuracy
chiral EFT calculations beyond the 2N system. 

\bigskip
\noindent
{\bf Acknowledgments}\\
The authors are grateful to Vadim Baru, Arseniy Filin, Ashot Gasparyan, Jambul
Gegelia, Christopher K\"orber, Ulf Mei{\ss}ner and to all members of the LENPIC Collaboration for sharing
their insights into the topics addressed in this chapter.
 This work was
   supported by BMBF (contract No. 05P18PCFP1), by
 DFG and NSFC through funds provided to the Sino-German CRC 110
 ``Symmetries and the Emergence of Structure in QCD'' (NSFC Grant
 No. 11621131001, DFG Project-ID 196253076 - TRR 110), by ERC AdG
 NuclearTheory (grant No. 885150) and by the EU Horizon 2020 research and
 innovation programme (STRONG-2020, grant agreement No. 824093).


\begin{thebibliography}{99.}

\bibitem{Reinert:2017usi}
P.~Reinert, H.~Krebs and E.~Epelbaum,
%``Semilocal momentum-space regularized chiral two-nucleon potentials up to fifth order,''
Eur. Phys. J. A \textbf{54}, no.5, 86 (2018).
%doi:10.1140/epja/i2018-12516-4
%[arXiv:1711.08821 [nucl-th]].

\bibitem{Reinert:2020mcu}
P.~Reinert, H.~Krebs and E.~Epelbaum,
%``Precision determination of pion-nucleon coupling constants using effective field theory,''
Phys. Rev. Lett. \textbf{126}, no.9, 092501 (2021).
%doi:10.1103/PhysRevLett.126.092501
%[arXiv:2006.15360 [nucl-th]].

\bibitem{Epelbaum:2014efa}
E.~Epelbaum, H.~Krebs and U.-G.~Mei\ss{}ner,
%``Improved chiral nucleon-nucleon potential up to next-to-next-to-next-to-leading order,''
Eur. Phys. J. A \textbf{51}, no.5, 53 (2015).
%doi:10.1140/epja/i2015-15053-8
%[arXiv:1412.0142 [nucl-th]].

\bibitem{Furnstahl:2015rha}
  R.~J.~Furnstahl {\it et al.},
  %, N.~Klco, D.~R.~Phillips and S.~Wesolowski,
%``Quantifying truncation errors in effective field theory,''
Phys. Rev. C \textbf{92}, no.2, 024005 (2015).
%doi:10.1103/PhysRevC.92.024005
%[arXiv:1506.01343 [nucl-th]].

\bibitem{Epelbaum:2019wvf}
E.~Epelbaum,
%``High-precision nuclear forces : Where do we stand?,''
PoS \textbf{CD2018}, 006 (2019).
%doi:10.22323/1.317.0006

\bibitem{Weinberg:1991um}
S.~Weinberg,
%``Effective chiral Lagrangians for nucleon - pion interactions and nuclear forces,''
Nucl. Phys. B \textbf{363}, 3-18 (1991).
%doi:10.1016/0550-3213(91)90231-L

\bibitem{Epelbaum:2019kcf}
E.~Epelbaum, H.~Krebs and P.~Reinert,
%``High-precision nuclear forces from chiral EFT: State-of-the-art, challenges and outlook,''
Front. in Phys. \textbf{8}, 98 (2020).
%doi:10.3389/fphy.2020.00098
%[arXiv:1911.11875 [nucl-th]].

\bibitem{Epelbaum:2009sd}
E.~Epelbaum and J.~Gegelia,
%``Regularization, renormalization and 'peratization' in effective field theory for two nucleons,''
Eur. Phys. J. A \textbf{41}, 341-354 (2009).
%doi:10.1140/epja/i2009-10833-3
%[arXiv:0906.3822 [nucl-th]].

\bibitem{Epelbaum:2018zli}
E.~Epelbaum, A.~M.~Gasparyan, J.~Gegelia and U.-G.~Mei\ss{}ner,
%``How (not) to renormalize integral equations with singular potentials in effective field theory,''
Eur. Phys. J. A \textbf{54},
    %     no.11,
186 (2018).
%doi:10.1140/epja/i2018-12632-1
%[arXiv:1810.02646 [nucl-th]].

\bibitem{Epelbaum:2019zqc}
  E.~Epelbaum {\it et al.},
  %, J.~Golak, K.~Hebeler, H.~Kamada, H.~Krebs, U.~G.~Mei\ss{}ner, A.~Nogga, P.~Reinert, R.~Skibi\'nski and K.~Topolnicki, \textit{et al.}
%``Towards high-order calculations of three-nucleon scattering in chiral effective field theory,''
Eur. Phys. J. A \textbf{56}, no.3, 92 (2020).
%doi:10.1140/epja/s10050-020-00102-2
%[arXiv:1907.03608 [nucl-th]].

%\bibitem{Kalantar-Nayestanaki:2011rzs}
%N.~Kalantar-Nayestanaki, E.~Epelbaum, J.~G.~Messchendorp and A.~Nogga,
%%``Signatures of three-nucleon interactions in few-nucleon systems,''
%Rept. Prog. Phys. \textbf{75}, 016301 (2012).
%%doi:10.1088/0034-4885/75/1/016301
%%[arXiv:1108.1227 [nucl-th]].

\bibitem{Hoferichter:2015tha}
  M.~Hoferichter {\it et al.},
  %, J.~Ruiz de Elvira, B.~Kubis and U.-G.~Mei\ss{}ner,
%``Matching pion-nucleon Roy-Steiner equations to chiral perturbation theory,''
Phys. Rev. Lett. \textbf{115}, no.19, 192301 (2015).
%doi:10.1103/PhysRevLett.115.192301
%[arXiv:1507.07552 [nucl-th]].

\bibitem{Girlanda:2020pqn}
L.~Girlanda, A.~Kievsky, L.~E.~Marcucci and M.~Viviani,
%``Unitary ambiguity of NN contact interactions and the 3N force,''
Phys. Rev. C \textbf{102}, 064003 (2020).
%doi:10.1103/PhysRevC.102.064003
%[arXiv:2007.04161 [nucl-th]].

\bibitem{Ekstrom:2015rta}
  A.~Ekstr\"om {\it et al.},
%  G.~R.~Jansen, K.~A.~Wendt, G.~Hagen, T.~Papenbrock, B.~D.~Carlsson, C.~Forss\'en, M.~Hjorth-Jensen, P.~Navr\'atil and W.~Nazarewicz,
%``Accurate nuclear radii and binding energies from a chiral interaction,''
Phys. Rev. C \textbf{91}, no.5, 051301 (2015).
%doi:10.1103/PhysRevC.91.051301
%[arXiv:1502.04682 [nucl-th]].

\bibitem{Epelbaum:2014sza}
E.~Epelbaum, H.~Krebs and U.-G.~Mei\ss{}ner,
%``Precision nucleon-nucleon potential at fifth order in the chiral expansion,''
Phys. Rev. Lett. \textbf{115}, no.12, 122301 (2015).
%doi:10.1103/PhysRevLett.115.122301
%[arXiv:1412.4623 [nucl-th]].

  \bibitem{Gasparyan:2021edy}
A.~M.~Gasparyan and E.~Epelbaum,
%``Nucleon-nucleon interaction in chiral EFT with a finite cutoff: explicit perturbative renormalization at next-to-leading order,''
[arXiv:2110.15302 [nucl-th]].

\bibitem{Gezerlis:2013ipa}
  A.~Gezerlis {\it et al.},
  %, I.~Tews, E.~Epelbaum, S.~Gandolfi, K.~Hebeler, A.~Nogga and A.~Schwenk,
%``Quantum Monte Carlo Calculations with Chiral Effective Field Theory Interactions,''
Phys. Rev. Lett. \textbf{111}, no.3, 032501 (2013).
%doi:10.1103/PhysRevLett.111.032501
%[arXiv:1303.6243 [nucl-th]].

\bibitem{Kaiser:1997mw}
N.~Kaiser, R.~Brockmann and W.~Weise,
%``Peripheral nucleon-nucleon phase shifts and chiral symmetry,''
Nucl. Phys. A \textbf{625}, 758-788 (1997).
%doi:10.1016/S0375-9474(97)00586-1
%[arXiv:nucl-th/9706045 [nucl-th]].

\bibitem{VanDerLeun:1982bhg}
C.~Van Der Leun and C.~Alderliesten,
%``The deuteron binding energy,''
Nucl. Phys. A \textbf{380}, 261-269 (1982).
%doi:10.1016/0375-9474(82)90105-1

\bibitem{Schoen:2003my}
  K.~Schoen {\it et al.},
  %, D.~L.~Jacobson, M.~Arif, P.~R.~Huffman, T.~C.~Black, W.~M.~Snow, S.~K.~Lamoreaux, H.~Kaiser and S.~A.~Werner,
%``Precision neutron interferometric measurements and updated evaluations of the n p and n d coherent neutron scattering lengths,''
Phys. Rev. C \textbf{67}, 044005 (2003).
%doi:10.1103/PhysRevC.67.044005

\bibitem{Stoks:1993tb}
  V.~G.~J.~Stoks {\it et al.},
  %, R.~A.~M.~Klomp, M.~C.~M.~Rentmeester and J.~J.~de Swart,
%``Partial wave analaysis of all nucleon-nucleon scattering data below 350-MeV,''
Phys. Rev. C \textbf{48}, 792-815 (1993).
%doi:10.1103/PhysRevC.48.792

\bibitem{NavarroPerez:2013mvd}
R.~Navarro P\'erez, J.~E.~Amaro and E.~Ruiz Arriola,
%``Coarse-grained potential analysis of neutron-proton and proton-proton scattering below the pion production threshold,''
Phys. Rev. C \textbf{88}, no.6, 064002 (2013)
[erratum: Phys. Rev. C \textbf{91}, no.2, 029901 (2015)].
%doi:10.1103/PhysRevC.88.064002
%[arXiv:1310.2536 [nucl-th]].

\bibitem{Reinert:2022thesis}
P.~Reinert,
%``Precision studies in the two-nucleon system using chiral effective field theory,''
PhD thesis, Ruhr-Universit\"at Bochum, to be published.

\bibitem{NavarroPerez:2016eli}
R.~Navarro P\'erez, J.~E.~Amaro and E.~Ruiz Arriola,
%``Precise Determination of Charge Dependent Pion-Nucleon-Nucleon Coupling Constants,''
Phys. Rev. C \textbf{95}, no.6, 064001 (2017).
%doi:10.1103/PhysRevC.95.064001
%[arXiv:1606.00592 [nucl-th]].

\bibitem{Gross:2008ps}
F.~Gross and A.~Stadler,
%``Covariant spectator theory of np scattering: Phase shifts obtained from precision fits to data below 350-MeV,''
Phys. Rev. C \textbf{78}, 014005 (2008).
%doi:10.1103/PhysRevC.78.014005
%[arXiv:0802.1552 [nucl-th]].
%127 citations counted in INSPIRE as of 26 Jan 2022

\bibitem{NavarroPerez:2013usk}
R.~Navarro P\'erez, J.~E.~Amaro and E.~Ruiz Arriola,
%``Partial Wave Analysis of Nucleon-Nucleon Scattering below pion production threshold,''
Phys. Rev. C \textbf{88}, 024002 (2013)
[erratum: Phys. Rev. C \textbf{88}, no.6, 069902 (2013)].
%doi:10.1103/PhysRevC.88.024002
%[arXiv:1304.0895 [nucl-th]].

\bibitem{Entem:2017gor}
D.~R.~Entem, R.~Machleidt and Y.~Nosyk,
%``High-quality two-nucleon potentials up to fifth order of the chiral expansion,''
Phys. Rev. C \textbf{96}, no.2, 024004 (2017).
%doi:10.1103/PhysRevC.96.024004
%[arXiv:1703.05454 [nucl-th]].
%206 citations counted in INSPIRE as of 27 Jan 2022

\bibitem{Machleidt:2000ge}
R.~Machleidt,
%``The High precision, charge dependent Bonn nucleon-nucleon potential (CD-Bonn),''
Phys. Rev. C \textbf{63}, 024001 (2001).
%doi:10.1103/PhysRevC.63.024001
%[arXiv:nucl-th/0006014 [nucl-th]].

\bibitem{Stoks:1994wp}
  V.~G.~J.~Stoks {\it et al.},
%  , R.~A.~M.~Klomp, C.~P.~F.~Terheggen and J.~J.~de Swart,
%``Construction of high quality N N potential models,''
Phys. Rev. C \textbf{49}, 2950-2962 (1994).
%doi:10.1103/PhysRevC.49.2950
%[arXiv:nucl-th/9406039 [nucl-th]].

\bibitem{Cox:1968jxz}
  G.~F.~Cox {\it et al.},
  %, G.~H.~Eaton, C.~P.~Van Zyl, O.~N.~Jarvis and B.~Rose,
%``Measurements of the differential cross section and polarization in proton-proton scattering at about 143 MeV,''
Nucl. Phys. B \textbf{4}, 353-373 (1968).
%doi:10.1016/0550-3213(68)90115-6

\bibitem{Jarvis:1971fla}
O.~N.~Jarvis, C.~Whitehead and M.~Shah,
%``Small-angle proton - proton scattering cross-sections at 144 MeV,''
Phys. Lett. B \textbf{36}, no.4, 409-411 (1971).
%doi:10.1016/0370-2693(71)90737-4

\bibitem{Taylor:1960}
A.~E.~Taylor, E.~Wood and L.~Bird,
%``Proton-proton scattering at 98 and 142 MeV,''
Nucl. Phys. \textbf{16}, no. 2, 320-330 (1960).
%doi:10.1016/S0029-5582(60)81041-3

\bibitem{Bird:1961}
L.~Bird, P.~Christmas, A.~Taylor and E.~Wood,
%``De-polarization parameter in p-p scattering at 143 MeV,''
Nucl. Phys. {\bf 27}, 586 (1961).
%doi:10.1016/0029-5582(61)90303-0



\bibitem{Lisowski:1982rm}
  P.~W.~Lisowski {\it et al.},
  %, R.~E.~Shamu, G.~F.~Auchampaugh, N.~S.~P.~King, M.~S.~Moore, G.~L.~Morgan and T.~S.~Singleton,
%``SEARCH FOR RESONANCE STRUCTURE IN N P TOTAL CROSS-SECTION BELOW 800-MeV,''
Phys. Rev. Lett. \textbf{49}, 255-259 (1982).
%doi:10.1103/PhysRevLett.49.255
%89 citations counted in INSPIRE as of 29 Jan 2022

\bibitem{Kessler:1999zz}
E.~G.~Kessler, Jr.,
%``The Deuteron Binding Energy and the Neutron Mass,''
Phys. Lett. A \textbf{255}, 221 (1999).
%doi:10.1016/S0375-9601(99)00078-X

\bibitem{deSwart:1995ui}
J.~J.~de Swart, C.~P.~F.~Terheggen and V.~G.~J.~Stoks,
%``The Low-energy n p scattering parameters and the deuteron,''
[arXiv:nucl-th/9509032 [nucl-th]].

\bibitem{Rodning:1990zz}
N.~L.~Rodning and L.~D.~Knutson,
%``Asymptotic D-state to S-state ratio of the deuteron,''
Phys. Rev. C \textbf{41}, 898-909 (1990).
%doi:10.1103/PhysRevC.41.898

\bibitem{Filin:2019eoe}
  A.~A.~Filin \textit{et al.},
  %V.~Baru, E.~Epelbaum, H.~Krebs, D.~M\"oller and P.~Reinert,
%``Extraction of the neutron charge radius from a precision calculation of the deuteron structure radius,''
Phys. Rev. Lett. \textbf{124}, no.8, 082501 (2020).
%doi:10.1103/PhysRevLett.124.082501
%[arXiv:1911.04877 [nucl-th]].  

\bibitem{Filin:2020tcs}
  A.~A.~Filin \textit{et al.},
  %D.~M\"oller, V.~Baru, E.~Epelbaum, H.~Krebs and P.~Reinert,
%``High-accuracy calculation of the deuteron charge and quadrupole form factors in chiral effective field theory,''
Phys. Rev. C \textbf{103}, no.2, 024313 (2021).
%doi:10.1103/PhysRevC.103.024313
%[arXiv:2009.08911 [nucl-th]].

\bibitem{Epelbaum:2022fjc}
  E.~Epelbaum {\it et al.},
  %J.~Gegelia, N.~Lange, U.-G.~Mei\ss{}ner and M.~V.~Polyakov,
%``On the definition of local spatial densities in hadrons,''
[arXiv:2201.02565 [hep-ph]].
  
\bibitem{Jentschura:2011}
  U.~D.~Jentschura {\it et al.},
  %, A.~Matveev, C.~G.~Parthey, J.~Alnis, R.~Pohl, Th.~Udem, N.~Kolachevsky and  T.~W.~HÃ¤nsch,
%``Hydrogen-deuterium isotope shift: From the1S-2S-transition frequency to the proton-deuteron charge-radius difference,''
Phys. Rev. A \textbf{83}, 042505 (2011).

\bibitem{Puchalski:2020jkt}
M.~Puchalski, J.~Komasa and K.~Pachucki,
%``Hyperfine Structure of the First Rotational Level in $H_2$, $D_2$ and HD Molecules and the Deuteron Quadrupole Moment,''
Phys. Rev. Lett. \textbf{125}, no.25, 253001 (2020).
%doi:10.1103/PhysRevLett.125.253001
%[arXiv:2010.06888 [physics.chem-ph]].

\bibitem{Maris:2020qne}
  P.~Maris {\it et al.},
  %, E.~Epelbaum, R.~J.~Furnstahl, J.~Golak, K.~Hebeler, T.~H\"uther, H.~Kamada, H.~Krebs, U.~G.~Mei\ss{}ner and J.~A.~Melendez, \textit{et al.}
%``Light nuclei with semilocal momentum-space regularized chiral interactions up to third order,''
Phys. Rev. C \textbf{103}, no.5, 054001 (2021).
%doi:10.1103/PhysRevC.103.054001
%[arXiv:2012.12396 [nucl-th]].


\bibitem{Hebeler:2015wxa}
  K.~Hebeler {\it et al.},
  %, H.~Krebs, E.~Epelbaum, J.~Golak and R.~Skibinski,
%``Efficient calculation of chiral three-nucleon forces up to N3LO for ab initio studies,''
Phys. Rev. C \textbf{91}, no.4, 044001 (2015).
%doi:10.1103/PhysRevC.91.044001
%[arXiv:1502.02977 [nucl-th]].

\bibitem{LENPIC:2018ewt}
E.~Epelbaum \textit{et al.} [LENPIC],
%``Few- and many-nucleon systems with semilocal coordinate-space regularized chiral two- and three-body forces,''
Phys. Rev. C \textbf{99}, no.2, 024313 (2019).
%doi:10.1103/PhysRevC.99.024313
%[arXiv:1807.02848 [nucl-th]].


\bibitem{Wesolowski:2021cni}
  S.~Wesolowski {\it et al.},
  %, I.~Svensson, A.~Ekstr\"om, C.~Forss\'en, R.~J.~Furnstahl, J.~A.~Melendez and D.~R.~Phillips,
%``Rigorous constraints on three-nucleon forces in chiral effective field theory from fast and accurate calculations of few-body observables,''
Phys. Rev. C \textbf{104}, no.6, 064001 (2021).
%doi:10.1103/PhysRevC.104.064001
%[arXiv:2104.04441 [nucl-th]].

\bibitem{Sekiguchi:2002sf}
  K.~Sekiguchi {\it et al.},
  %, H.~Sakai, H.~Witaa, W.~Gloeckle, J.~Golak, M.~Hatano, H.~Kamada, H.~Kato, Y.~Maeda and J.~Nishikawa, \textit{et al.}
%``Complete set of precise deuteron analyzing powers at intermediate energies: comparison with modern nuclear force predictions,''
Phys. Rev. C \textbf{65}, 034003 (2002).
%doi:10.1103/PhysRevC.65.034003

\bibitem{Gloeckle:1995jg}
W.~Gl{\"o}ckle, H.~Wita{\l}a, D.~H\"uber, H.~Kamada and J.~Golak,
%``The Three nucleon continuum: Achievements, challenges and applications,''
Phys. Rept. \textbf{274}, 107-285 (1996).
%doi:10.1016/0370-1573(95)00085-2
  
\bibitem{Abfalterer:1998zz}
  W.~P.~Abfalterer {\it et al.},
  %, F.~B.~Bateman, F.~S.~Dietrich, C.~Elster, R.~W.~Finlay, W.~Glockle, J.~Golak, R.~C.~Haight, D.~Huber and G.~L.~Morgan, \textit{et al.}
%``Inadequacies of the nonrelativistic 3N Hamiltonian in describing the n + d total cross section,''
Phys. Rev. Lett. \textbf{81}, 57-60 (1998).
%doi:10.1103/PhysRevLett.81.57

\bibitem{LENPIC:2015qsz}
S.~Binder \textit{et al.} [LENPIC],
%``Few-nucleon systems with state-of-the-art chiral nucleon-nucleon forces,''
Phys. Rev. C \textbf{93}, no.4, 044002 (2016).
%doi:10.1103/PhysRevC.93.044002
%[arXiv:1505.07218 [nucl-th]].

\bibitem{LENPIC:2018lzt}
S.~Binder \textit{et al.} [LENPIC],
%``Few-nucleon and many-nucleon systems with semilocal coordinate-space regularized chiral nucleon-nucleon forces,''
Phys. Rev. C \textbf{98}, no.1, 014002 (2018).
%doi:10.1103/PhysRevC.98.014002
%[arXiv:1802.08584 [nucl-th]].

  
\bibitem{Bernard:2011zr} 
  V.~Bernard, E.~Epelbaum, H.~Krebs and U.-G.~Mei{\ss}ner,
  %``Subleading contributions to the chiral three-nucleon force II: Short-range terms and relativistic corrections,''
  Phys.\ Rev.\ C {\bf 84}, 054001 (2011).
%  doi:10.1103/PhysRevC.84.054001
%  [arXiv:1108.3816 [nucl-th]].
  %%CITATION = doi:10.1103/PhysRevC.84.054001;%%
  %130 citations counted in INSPIRE as of 25 Sep 2019

%\cite{Krebs:2020pii}
\bibitem{Bernard:2007sp}
V.~Bernard, E.~Epelbaum, H.~Krebs and U.-G.~Mei{\ss}ner,
%``Subleading contributions to the chiral three-nucleon force. I. Long-range terms,''
Phys. Rev. C \textbf{77}, 064004 (2008).
%doi:10.1103/PhysRevC.77.064004
%[arXiv:0712.1967 [nucl-th]].

\bibitem{Krebs:2020pii}
H.~Krebs,
%``Nuclear Currents in Chiral Effective Field Theory,''
Eur. Phys. J. A \textbf{56}, no.9, 234 (2020).
%doi:10.1140/epja/s10050-020-00230-9
%[arXiv:2008.00974 [nucl-th]].
%11 citations counted in INSPIRE as of 27 Jan 2022

%\cite{Slavnov:1971aw}
\bibitem{Slavnov:1971aw}
A.~A.~Slavnov,
%``Invariant regularization of nonlinear chiral theories,''
Nucl. Phys. B \textbf{31}, 301-315 (1971).
%doi:10.1016/0550-3213(71)90234-3
%160 citations counted in INSPIRE as of 27 Jan 2022

%\cite{Luscher:2010iy}
\bibitem{Luscher:2010iy}
M.~L\"uscher,
%``Properties and uses of the Wilson flow in lattice QCD,''
JHEP \textbf{08}, 071 (2010)
[erratum: JHEP \textbf{03}, 092 (2014)].
%doi:10.1007/JHEP08(2010)071
%[arXiv:1006.4518 [hep-lat]].
%801 citations counted in INSPIRE as of 27 Jan 2022

%\cite{Grabowska:2015qpk}
\bibitem{Grabowska:2015qpk}
D.~M.~Grabowska and D.~B.~Kaplan,
%``Nonperturbative Regulator for Chiral Gauge Theories?,''
Phys. Rev. Lett. \textbf{116}, no.21, 211602 (2016).
%doi:10.1103/PhysRevLett.116.211602
%[arXiv:1511.03649 [hep-lat]].
%36 citations counted in INSPIRE as of 27 Jan 2022

%\cite{Kaplan:2016dgz}
\bibitem{Kaplan:2016dgz}
D.~B.~Kaplan and D.~M.~Grabowska,
%``A New Perspective on Chiral Gauge Theories,''
PoS \textbf{LATTICE2016}, 018 (2016).
%doi:10.22323/1.256.0018
%3 citations counted in INSPIRE as of 27 Jan 2022

%\cite{Bar:2013ora}
\bibitem{Bar:2013ora}
O.~B\"ar and M.~Golterman,
%``Chiral perturbation theory for gradient flow observables,''
Phys. Rev. D \textbf{89}, no.3, 034505 (2014)
[erratum: Phys. Rev. D \textbf{89}, no.9, 099905 (2014)].
%doi:10.1103/PhysRevD.89.034505
%[arXiv:1312.4999 [hep-lat]].
%73 citations counted in INSPIRE as of 27 Jan 2022

%\cite{Aoki:2014dxa}
\bibitem{Aoki:2014dxa}
S.~Aoki, K.~Kikuchi and T.~Onogi,
%``Gradient Flow of O(N) nonlinear sigma model at large N,''
JHEP \textbf{04}, 156 (2015).
%doi:10.1007/JHEP04(2015)156
%[arXiv:1412.8249 [hep-th]].
%21 citations counted in INSPIRE as of 27 Jan 2022

%\cite{Borasoy:2003pg}
\bibitem{Borasoy:2003pg}
B.~Borasoy, R.~Lewis and P.~P.~A.~Ouimet,
%``Lattice regularized chiral perturbation theory,''
Nucl. Phys. B Proc. Suppl. \textbf{128}, 141-147 (2004).
%doi:10.1016/S0920-5632(03)02470-8
%[arXiv:hep-lat/0310054 [hep-lat]].
%9 citations counted in INSPIRE as of 27 Jan 2022

%\cite{Lee:2020meg}
\bibitem{Lahde:2019npb}
T.~A.~L\"ahde and U.-G.~Mei\ss{}ner,
%``Nuclear Lattice Effective Field Theory: An introduction,''
Lect. Notes Phys. \textbf{957}, 1-396 (2019).
%doi:10.1007/978-3-030-14189-9


\bibitem{Lee:2020meg}
D.~Lee,
%``Recent Progress in Nuclear Lattice Simulations,''
Front. in Phys. \textbf{8}, 174 (2020).
%doi:10.3389/fphy.2020.00174
%5 citations counted in INSPIRE as of 27 Jan 2022



\bibitem{Epelbaum:2005bjv}
E.~Epelbaum,
%``Four-nucleon force in chiral effective field theory,''
Phys. Lett. B \textbf{639}, 456-461 (2006).
%doi:10.1016/j.physletb.2006.06.046
%[arXiv:nucl-th/0511025 [nucl-th]].

\bibitem{Epelbaum:2007us}
E.~Epelbaum,
%``Four-nucleon force using the method of unitary transformation,''
Eur. Phys. J. A \textbf{34}, 197-214 (2007).
%doi:10.1140/epja/i2007-10496-0
%[arXiv:0710.4250 [nucl-th]].

\bibitem{Girlanda:2011fh}
L.~Girlanda, A.~Kievsky and M.~Viviani,
%``Subleading contributions to the three-nucleon contact interaction,''
Phys. Rev. C \textbf{84}, no.1, 014001 (2011)
[erratum: Phys. Rev. C \textbf{102}, no.1, 019903 (2020)].
%doi:10.1103/PhysRevC.84.014001
%[arXiv:1102.4799 [nucl-th]].

\bibitem{Krebs:2012yv}
H.~Krebs, A.~Gasparyan and E.~Epelbaum,
%``Chiral three-nucleon force at N$^4$LO I: Longest-range contributions,''
Phys. Rev. C \textbf{85}, 054006 (2012).
%doi:10.1103/PhysRevC.85.054006
%[arXiv:1203.0067 [nucl-th]].

\bibitem{Krebs:2013kha}
H.~Krebs, A.~Gasparyan and E.~Epelbaum,
%``Chiral three-nucleon force at $N^4LO$ II: Intermediate-range contributions,''
Phys. Rev. C \textbf{87}, no.5, 054007 (2013).
%doi:10.1103/PhysRevC.87.054007
%[arXiv:1302.2872 [nucl-th]].

\end{thebibliography}
\end{document}